\documentclass[11pt]{article}

\usepackage{amsmath,amssymb}
\newtheorem{definition}{\noindent \noindent {\bf Definition}}[section]
\newtheorem{lemma}{{\bf Lemma}}[section]
\newtheorem{theorem}{Theorem}[section]
\newtheorem{proposition}{{\bf Proposition}}[section]
\newtheorem{corollary}{{\bf Corollary}}[section]
\newtheorem{remark}{{\bf Remark}}[section]

\newtheorem{state}{{\bf Statement}}[section]

 1   %Caracteres goticos.
 1  %Caracteres "doble palo".

\newenvironment{proof}{\noindent{\it Proof:~}}{\qed\medskip}
\def\QED{\hskip0.1em\hfill\null\ \null\nobreak\hfill
\kern3pt\lower1.8pt\vbox{\hrule\hbox
{\vrule\kern1pt\vbox{\kern1.7pt \hbox{$\scriptstyle
QED$}\kern0.2pt}\kern1pt\vrule}\hrule}}
 \def\hook{\, \hbox to 15pt{\vbox{\vskip 6pt\hrule width 8pt height 1pt}
         \kern -5pt\vrule height 8pt width 1pt\hfil}}
\def\blob{\quad\rule{8pt}{8pt}}

\def\r{\ensuremath{\mathbb{R}}}
\def\rk{{\mathbb R}^{k}}

\def\K{{\mathcal{K}}}
\newcommand{\ds}{\displaystyle}

\def\feble#1{\mathrel{\mathop =\limits_{#1}}}
\def\Cinfty{{\rm C}^\infty}

\begin{document}

\parskip=8pt

\begin{center}
{\Large \bf\sf    G\"unther's formalism ($k$-symplectic
formalism) in classical field theory: Skinner--Rusk approach and
the evolution operator}
\end{center}

\noindent
    Angel M. Rey  \\
    {\it Departamento de Xeometr\'{\i}a e Topolox\'{\i}a,
    Facultade de Matem\'{a}ticas,\\
    Universidade de Santia\-go de Compostela,
    15706-Santia\-go de Compostela, Spain} \\
    e-mail: angelmrey@edu.xunta.es\\

\noindent
    Narciso Rom\'an-Roy\\
    {\it Departamento deMatem\'{a}tica Aplicada IV, Edificio C-3, Campus Norte
    UPC,\\
    C/ Jordi Girona 1, E-08034 Barcelona, Spain}\\
    e-mail: matnrr@mat.upc.es\\

\noindent
    Modesto  Salgado\\
    {\it Departamento de Xeometr\'{\i}a e Topolox\'{\i}a,
    Facultade de Matem\'{a}ticas,\\
    Universidade de Santia\-go de Compostela,
    15706-Santia\-go de Compostela, Spain }\\
    e-mail:  modesto@zmat.usc.es

\bigskip

%\today

\bigskip

\bigskip

\bigskip

\begin{abstract}
The first aim of this paper is to extend the Skinner-Rusk
formalism on classical mechanics  for first-order field theories.
 The second is to generalize
the definition and properties of the evolution $K$-operator on
classical mechanics   for first-order field theories using in both
cases G\"unther's formalism ($k$-symplectic formalism).
\end{abstract}

\noindent M.S. Classification (2000): 70S05, 53D05, 53Z05

\renewcommand{\baselinestretch}{1.5}
\renewcommand{\arraystretch}{0.66}

%\normalsize

\clearpage

\section{Introduction}

The {\it Skinner-Rusk  formalism} \cite{skinner2}
 was developed in order to give a
geometrical unified formalism for describing mechanical systems.
It incorporates all the characteristics of Lagrangian and
Hamiltonian descriptions of these systems (including dynamical
equations and solutions, constraints, Legendre map, evolution
operators, equivalence, etc.).

This formalism has been generalized to time-dependent mechanical
systems \cite{CMC}, and also to the multisymplectic description of
first-order field theories \cite{bar4} , \cite{LMM-2002}.

The first aim of this paper is to extend this unified framework to
G\"unther's description of first-order classical field
theories \cite{gun}, and show how this description comprises the
main features of the Lagrangian and Hamiltonian formalisms, both
for the regular and singular cases.

Let us point out that G\"unther's formalism should be also
called $k$-symplectic formalism because the base of this formalism
are the standard polysymplectic manifolds, introduced by G\"unther
in \cite{gun}, which coincide with the $k$-symplectic manifolds
introduced by Awane in \cite{aw1,aw2,aw3}. G\"unther's paper gives
a geometric Hamiltonian formalism for field theories. The crucial
device is the introduction of a vector-valued generalization of a
symplectic form, called a polysymplectic form. One of the
advantages of this formalism is that only the tangent
and cotangent bundle of a manifold are required to develop it. In \cite{fam}
G\"unther's formalism was revised and clarified. It was
shown that the polysymplectic sructures used by G\"unther to
develop his formalism could  be replaced by the $k$-symplectic
structures defined by Awane \cite{aw1,aw2,aw3}. So this formalism
could be called $k$-symplectic formalism.

The $k$-symplectic formalism is the generalization to field
theories of the standard symplectic formalism in mechanics, which
is the geometric framework for describing autonomous dynamical
systems. In this sense, the $k$-symplectic formalism is used to
give a geometric description of certain kind of field theories: in
a local description, those whose Lagrangian does not depend on the
coordinates in the basis (in many of them, the space-time
coordinates); that is, it is only valid for Lagrangian
$L(q^i,v^i_A)$  and  Hamiltonian $H(q^i,p^A_i)$ that depends on
the field coordinates $q^i$ and  on the partial derivatives of the
field $v^i_A$. A natural extension of this formalism is the
so-called $k$-cosymplectic formalism, which is the generalization
to field theories of the cosymplectic formalism which describes
geometrically non-autonomous mechanical systems (this description
can be found in \cite{mod1,mod2}). It is devoted to describing field
theories involving the independent parameters $(t^1,\ldots ,t^k)$
on the Lagrangian $L(t^A,q^i,v^i_A)$ and on the Hamiltonian
$H(t^A,q^i,p^A_i)$.

It is interesting to remark here that the polysymplectic formalism
developed by G. Sardanashvily {\it et al}
\cite{Sarda2,Sarda1,Sd-95}, based on a vector valued form on some
associated fiber  bundle, is a different description of classical
field theories of first order than the polysymplectic formalism
proposed by G\"unther. (See also \cite{Kana} for more details on
the polysymplectic formalism).
  In addition, we must remark   that the soldering form on the linear
frames   bundles is a polysymplectic form, and its study and
applications to field theory constitute the $n$-symplectic
geometry developped by L. K. Norris in \cite{No2,No3,No4,No5,McN}.

The so-called {\sl time-evolution $K$-operator} in mechanics (also
known by some authors as the {\sl relative Hamiltonian vector
field} \cite{PV-00}) is a tool which has mainly been developed in
order to study the Lagrangian and Hamiltonian formalisms for
singular mechanical systems and their equivalence. This operator
was introduced in \cite{BGPR-eblhf} and \cite{Ka-82}, and later it
was defined geometrically in two different but equivalent ways
\cite{CL-92}, \cite{GP-01} for autonomous dynamical systems. In
\cite{GP-01}, a further different geometric construction is given,
using a canonical map introduced by Tulczyjew \cite{Tu-76}. The
$K$-operator relates the sets of solutions of the Euler-Lagrange
equations and the Hamilton equations; it also relates constraints
on the Lagrangian and Hamiltonian sides, and allows us to obtain a
complete classification of constraints \cite{BGPR-eblhf}; as well
as Lagrangian Noether infinitesimal symmetries from a Hamiltonian
generator of symmetries \cite{PV-00,FP-90,GaP-00,GP-92b}. It is
also used for studying Lagrangian systems whose Legendre map has
generic singularities \cite{PV-00,PV-00b}.

The second aim of this paper is to generalize the definition and
properties of this operator    for first-order field theories in
order to describe the relationship between the Lagrangian and
Hamiltonian $k$-symplectic formalisms. In particular we  extend
the results in \cite{GP-01}, showing how to obtain  the solutions
of Lagrangian and Hamiltonian field equations by means of this
operator. The same idea has been developed in \cite{bar5} but
using the multisymplectic description of classical field theories.

The organization of the paper is as follows: Section 2-4 are
devoted to reviewing the main features of G\"unther's
formalism or $k$-symplectic formalism \cite{gun,fam} of Lagrangian
and Hamiltonian field theories.

In particular, in Section 2 the field theoretic phase space is
introduced as the Whitney sum $(T^1_k)^*Q=T^*Q\oplus
\stackrel{k}{\dots} \oplus T^*Q$ of $k$-copies of the cotangent
bundle $T^*Q$ of a manifold $Q$. This space is the canonical
example of a polysymplectic manifold. A particular case of
polysymplectic manifolds are the $k$-symplectic manifolds (see
Refs. $[3,4,5,8,9]$) which coincide with the standard
polysymplectic manifolds.

The field theoretic state space is introduced as the Whitney sum
$  T^1_k Q=TQ\oplus \stackrel{k}{\dots} \oplus TQ$  of $k$-copies
of the tangent bundle $TQ$ of a manifold $Q$. This manifold has  a
canonical $k$-tangent structure defined by $k$ tensor fields of
type $(1,1)$ satisfying certain algebraic properties. The
$k$-tangent manifolds were introduced in de Le\'on {\it et al.}
\cite{mt1,mt2}, and they generalize the tangent manifolds (see
Refs. \cite{cgt,e,grif1,grif2,klein,mt2}).

Section 3 is devoted to giving a geometric interpretation of the
second order partial differential equations. Here we show that
these equations can be characterized by using the canonical
$k$-tangent structure of $T^1_kQ$, which generalizes the case of
Classical Mechanics.

The Hamiltonian and Lagrangian formalisms are developed in Section
4.  Lagrangian formalism is developed using the canonical
$k$-tangent structure of $T^1_kQ$,  or the Legendre transformation
as in G\"unther \cite{gun} .

In section 5 we develop the unified formalism for field theories,
which is based on the use of the Whitney sum $T^1_kQ\oplus_Q
(T^1_k)^*Q$  of $T^1_kQ$ and $(T^1_k)^*Q$.  There are canonical
presymplectic forms on it (the pull-back of the canonical
symplectic form on each $T^*Q$) and a natural coupling function
which is defined by the contraction between vectors and covectors.
Then, given a Lagrangian $L\in C^{\infty}(T^1_kQ)$ we can state a
field equation on $T^1_kQ\oplus_Q(T^1_k)^*Q$. This equation has
solution only on a submanifold $M_L$, which is the graph of the
Legendre map. Then we prove that if ${\bf Z}=(Z_1, \ldots , Z_k)$
is an integrable $k$-vector field,  solution to this equation and
tangent to $M_L$, then the projection onto the first factor
$T^1_kQ$ of the integral
 sections of ${\bf Z}$ are solutions of the Euler-Lagrange
 field equations. If $L$ is regular the converse also holds.
 Furthermore, we establish the relationship between ${\bf Z}$ and the
 Hamiltonian and the Lagrangian $k$-vector
 fields of the $k$-symplectic formalism, ${\bf X_H}$ and ${\bf X_L}$.

In Section 6 we review the definition and the main properties of
the evolution operator $K$ for autonomous mechanics. Next we
define the field operators which, as a consequence of the field
equations on the  $k$-symplectic formalism, are given as a
$k$-vector field along the Legendre transformation $FL$,
associated to the lagrangian $L:T^1_kQ \to \r$, satisfying certain
properties. Finally we finish with similar results for field
theories to those obtained in \cite{GP-01} and \cite{bar5}.

In a forthcoming paper we shall extend the results of this paper
to the $k$-cosymplectic formalism \cite{mod1,mod2}.

Manifolds are real, paracompact, connected and $C^\infty$. Maps
are $C^\infty$. Sum over crossed repeated indices is understood.

\section{Geometric framework: autonomous case}

\subsection{The cotangent bundle of $k^1$-covelocities of a manifold}

 Let $Q$ be a
differentiable manifold of dimension  $n$ and $\tau^*: T^*Q \to Q$
its cotangent bundle. Let us denote by $(T^1_k)^*Q= T^*Q \oplus
\stackrel{k}{\dots} \oplus T^*Q$ the Whitney sum of $k$ copies of
$T^*Q$, with projection map $\tau^*_Q\colon (T^1_k)^* Q \to Q$,
$\tau^*_Q (\alpha^1_q,\ldots ,\alpha^k_q)=q$, for every
$(\alpha^1_q,\ldots ,\alpha^k_q)\in (T^1_k)^*Q$.

$(T^1_k)^*Q$ can be canonically identified with the vector bundle
$J^1(Q,\r^k)_0$ of $k^1$-covelocities of the manifold $Q$,  that
is the vector bundle of $1$-jets of maps $\sigma\colon Q\to\rk$
with target at $0\in \r^k$ and projection map $\tau^*_Q :
J^1(Q,\r^k)_0  \to Q$, $\tau_Q^* (j^1_{q,0}\sigma)=q$, \, say ,
\[
\begin{array}{ccc}
J^1(Q,\r^k)_0 & \equiv & T^*Q \oplus \stackrel{k}{\dots} \oplus T^*Q \\
 j^1_{q,0}\sigma  & \equiv & (d\sigma^1(q), \dots ,d\sigma^k(q))
\end{array}
\]
where $\sigma^A= \pi^A \circ \sigma:Q \longrightarrow \r$ is the
 $A^{th}$ component of $\sigma$, and  $\pi^A:\r^k \to \r$ is the canonical projection $1\leq A \leq k$.
 For this reason to $(T^1_k)^*Q$ is also called {\it the bundle of
 $k^1$-covelocities of the manifold $Q$.}

If $(q^i)$ are local coordinates on $U \subseteq Q$, then the
induced local coordinates $(q^i , p_i)$, $1\leq i \leq n$, on
$T^*U=(\tau^*)^{-1}(U)$, are given by
$$
q^i( \alpha_q)=q^i(q), \quad p_i( \alpha_q)=
\alpha_q\left(\frac{\partial}{\partial q^i}\Big\vert_q
\right)\,,\quad \alpha_q \in T^*Q \,  ,
$$
and the induced local coordinates $(q^i , p^A_i),\, 1\leq i \leq
n,\, 1\leq A \leq k$, on $(T^1_k)^*U=(\tau_Q^*)^{-1}(U)$ are given
by
$$
q^i(\alpha^1_q,\ldots , \alpha^k_q)=q^i(q),\qquad
p^A_i(\alpha^1_q,\ldots ,\alpha^k_q )= \alpha^A_q
\left(\frac{\partial}{\partial q^i}\Big\vert_q\right) \quad .
$$

Let us denote by $\{r_1, \ldots, r_k\}$ the canonical basis of
 $\rk $.
\noindent \begin{definition}  (G\"unther \cite{gun}) A closed
non-degenerate $\r^k$-valued $2$-form
$$\bar{\omega} =\displaystyle \sum_{A=1}^k \omega_A \otimes r_A$$
on a manifold $M$ of dimension $N$ is called a {\rm polysymplectic
form}. The pair $(M,\bar{\omega})$ is a {\rm polysymplectic
manifold}.
\end{definition}

The manifold $(T^1_k)^*Q$ is endowed with a   {\it canonical
polysymplectic structure}. This canonical   structure
$\bar{\omega} = \sum_{A=1}^k (\omega_0)_A \otimes r_A$, on
$(T^1_k)^*Q$ is defined by
$$
(\omega_0)_A = (\tau^*_A)^*(\omega_0), \quad 1\leq A \leq k\, ,
$$
where $\tau^*_A : (T^1_k)^*Q \rightarrow T^*Q $ is the projection
on the $A^{th}$-copy  $T^*Q$ of $(T^1_k)^*Q$, and
$\omega_0=-d\theta_0$ is the canonical symplectic structure of
$T^*Q$, $\theta_0$ being the Liouville $1$-form defined by
$$\theta_0(\alpha_q)(\widetilde{X}_{\alpha_q})=\alpha_q((\tau^*)_*(\alpha_q)
(\widetilde{X}_{\alpha_q})), \quad \alpha_q\in T^*Q, \,\,
\widetilde{X}_{\alpha_q}\in T_{\alpha_q}(T^*Q).$$

One can also define the $2$-forms $(\omega_0)_A $ by $(\omega_0)_A
= -d(\theta_0)_A$
 where $(\theta_0)_A=(\tau_A^*)^*\theta_0$.

 Thus the Liouville $1$-form and the canonical symplectic structure
on $T^*Q$ are locally given by
$$
\theta_0=p_i\, dq^i, \qquad \omega_0= -d\theta_0= dq^i\wedge
dp_{i}
  \, ,$$
  and the canonical polysymplectic structure  $((\omega_0)_1, \ldots , (\omega_0)_k)$ on $(T^1_k)^*Q$ is locally given
by
\begin{equation}\label{canosym}
(\omega_0)_A= -d(\theta_0)_A=- d(p^A_i \, dq^i )=dq^i\wedge
dp^A_{i}
  \, .\end{equation}

 \begin{definition}  (G\"unther \cite{gun})
 A polysymplectic form $\bar{\omega}$ on
a manifold $M$ is called {\rm standard} iff for every point of $M$
there exists a local coordinate system such that $\omega_A$ is
written locally as in (\ref{canosym}).
\end{definition}

So the canonical polysymplectic form $\bar{\omega}$ on
$(T^1_k)^*Q$ is standard.

\begin{remark}{\rm
  The    $k$-symplectic manifolds were introduced
 in Awane
\cite{aw1,aw2,aw3} and they coincide with the {\it standard
polysymplectic} manifolds, as we now shall show. }
\end{remark}

 \begin{definition}\label{defaw} (Awane \cite{aw1}) A
$k$-symplectic structure on  a manifold $M$ of dimension $N=n+kn$
is a  family $(\omega_A,V;1\leq A\leq k)$, where each $\omega_A$
is a closed $2$-form and $V$ is an integrable $nk$-dimensional
distribution on $M$ such that
 $$(i) \quad
\omega_{A_{\vert V\times V}}=0,\qquad (ii) \quad \ds\cap_{A=1}^{k}
\ker\omega_A=\{0\}.$$ In this case $(M,\omega_A,V)$ is called a
$k$-symplectic manifold.
\end{definition}

\begin{theorem}(Awane \cite{aw1})
 Let $(\omega_A,V;1\leq A\leq k)$ be a $k$-symplectic
structure on $M$. About every point of $M$ we can find a local
coordinate system $(q^i , p^A_i),\, 1\leq i \leq n,\, 1\leq A \leq
k$, such that
\begin{equation}\label{locksym}
\omega_A=  dq^i\wedge dp^A_{i}, \quad  1\leq A \leq k \quad .
\end{equation}
\label{thks}
\end{theorem}

 The canonical model of $k$-symplectic
manifolds is  also $(T^1_k)^*Q$ and the canonical $k$-symplectic
structure $(\omega_A,V;1\leq A\leq k)$, on $(T^1_k)^*Q$ is given
by
$$
\omega_A = (\omega_0)_A=(\tau^*_A)^*(\omega_0), \qquad
V(j^1_{q,0}\sigma) = \ker(\tau^*_Q)_*(j^1_{q,0}\sigma)\quad .
$$

Therefore,  the $2$-forms of the canonical polysymplectic
structure and  the canonical $k$-symplectic structure on
$(T^1_k)^*Q$ coincide.

 From (\ref{locksym}) we know that  the standard polysymplectic
structures and the
    $k$-symplectic structures coincide. Indeed, if
$\bar{\omega} = \sum_{A=1}^k \omega_A \otimes r_A$ is a
    standard polysymplectic structure on $M$, given a local adapted
    coordinate system $(q^i,p^A_i)$ we can define,
 locally, the distribution $V$, of dimension  $nk$,
    by $dq^1= \ldots = dq^n=0$.
Then , $(\omega_1, \ldots ,\omega_k,V)$  is a $k$-symplectic structure on $M$.

    Conversely if $(\omega_1, \ldots ,\omega_k,V)$  is a $k$-symplectic
    structure on $M$ then $\bar{\omega} = \sum_{A=1}^k \omega_A \otimes r_A$ is a
    standard polysymplectic structure on $M$,
    because it is trivially standard and is non
    degenerate as a consequence of (ii) in Definition \ref{defaw}.

As we shall see later, in his Hamiltonian formalism,
G\"unther uses a standard polysymplectic manifold because he needs to
have local coordinates $(q^i,p^A_i)$ in the manifold $M$ where the Hamiltonian
is defined, which is equivalent to considering a
$k$-symplectic manifold. For this reason we will call the
G\"unther's formalism, called polysymplectic formalism,
{\ $k$-symplectic formalism}.

\subsection{The tangent bundle of $k^1$-velocities of a manifold}

Let $\tau : TQ \to Q$ be the tangent bundle of $Q$.
Let us denote by $T^1_kQ$
 the Whitney sum $TQ \oplus \stackrel{k}{\dots} \oplus TQ$ of $k$
copies of $TQ$, with projection $\tau_Q\colon T^1_kQ \to Q$,
$\tau_Q ({v_1}_q,\ldots , {v_k}_q)=q$.

$T^1_kQ$ can be identified with the vector bundle $J^1_0(\r^k,Q)$
of the $k^1$-velocities of the manifold $Q$, that is, the vector
bundle of   $1$-jets of maps $\sigma\colon\rk\to Q$  with source
at $0\in \r^k$, and
  projection map $\tau_Q : T^1_kQ \to Q$, $\tau_Q
(j^1_{0,q}\sigma)=\sigma (0)=q$, say
\[
\begin{array}{ccc}
J^1_0(\r^k,Q) & \equiv & TQ \oplus \stackrel{k}{\dots} \oplus TQ \\
 j^1_{0,q}\sigma  & \equiv & ({v_1}_q,\ldots , {v_k}_q)
\end{array}
\]
where $q=\sigma (0)$,  and ${v_A}_q=\sigma_*(0)[(\partial/\partial
t^A)(0)],\, 1\leq A\leq k$. For this reason $T^1_kQ$ is called the
{\it tangent bundle of $k^1$-velocities of $Q$.}

If $(q^i)$ are local coordinates on $U \subseteq Q$ then the
induced local coordinates $(q^i , v^i)$, $1\leq i \leq n$, on
$TU=\tau^{-1}(U)$ are given by
$$
q^i(v_q)=q^i(q),\qquad  v^i( v_q)=v_q(q^i),\qquad  v_q\in TQ \, ,
$$
and  the induced local coordinates $(q^i,v_A^i)$, $1\leq i \leq
n,\, 1\leq A \leq k$, on $T^1_kU=\tau_Q^{-1}(U)$ are given by
$$ q^i({v_1}_q,\ldots , {v_k}_q)=q^i(q),\qquad
  v_A^i({v_1}_q,\ldots , {v_k}_q)={v_A}_q(q^i) \, .$$

 We now introduce the {\it canonical $k$-tangent structure} on $T^1_kQ$.

\begin{definition} For a vector  $X_q$ at $Q$, and
for $1 \leq A \leq k$, we define its {\it vertical $A$-lift} \,
$(X_q)^A$ as the vector   on $T_k^1Q$ given by
  $$(X_q)^A({v_1}_q,\ldots , {v_k}_q) = \displaystyle\frac{d}{ds} (
{v_1}_q,\ldots,{v_{A-1}}_q,{v_A}_q+sX_q,{v_{A+1}}_q,
\ldots,{v_k}_q)_{\vert_{s=0}} $$ \noindent for all points
$({v_1}_q,\ldots , {v_k}_q)\in T^1_kQ$.
\end{definition}

In local coordinates we have
\begin{equation}\label{xa}
(X_q)^A =  a^i \displaystyle\frac{\partial}{\partial
v^i_A}\Big\vert_q
\end{equation}
for a vector $X_q = a^i \,(\partial/\partial q^i)(q)$.

The {\it canonical $k$-tangent structure} on $T^1_kQ$ is the set
$(S^1,\ldots,S^k)$ of tensor fields  of type $(1,1)$ defined by
 $$S^A(v)(Z_{v})=
  ((\tau_Q)_*(v)(Z_{v}))^A, \quad \mbox{for all} \, \, Z_{v}\in T_{v}(T^1_kQ), \, v=({v_1}_q,\ldots , {v_k}_q),$$
\noindent   for each $1 \leq A \leq k$.

  From (\ref{xa})  we have in
local  coordinates
\begin{equation}\label{localJA}
S^A=\displaystyle\frac
{\displaystyle\partial}{\displaystyle\partial v^i_A} \otimes dq^i
\end{equation}

  The tensors $S^A$ can be regarded as the
$(0,\ldots,0,\stackrel{A}{1},0,\ldots,0)$-lift of the identity
tensor on $Q$ to $T^1_kQ$ defined by Morimoto \cite{mor}.

\begin{remark} {\rm The $k$-tangent manifolds were introduced
as a generalization of the tangent manifolds by  de Le\'on {\it et
al.} \cite{mt1,mt2}. The canonical model of these manifolds is
$T^1_kQ$ with the
    structure given by $(S^1, \ldots , S^k)$.}
\end{remark}

 To develop later the Lagrangian formalism,
we now construct a polysymplectic structure on $T^1_kQ$, for each
regular Lagrangian $L:T^1_kQ\to \r$, , using its canonical
$k$--tangent structure.

\begin{definition} A Lagrangian $L:T^1_kQ
\rightarrow \r$  is called {\it regular} if and only if
$$
det \left(\frac{\displaystyle\partial^2L} {\displaystyle\partial
v^i_A
\partial v^j_B}\right)\neq 0,
\qquad   1 \leq i,j
 \leq n,\quad 1\leq A,B \leq k  \, .
$$
\end{definition}

Let us consider the $1$--forms $(\theta_L)_A = dL \circ S^A \, ,\,
1\leq A \leq k$. In a local coordinate system $(q^i,v^i_A)$ we
have
\begin{equation}\label{betaloc}
(\theta_L)_A=\frac{\displaystyle \partial
 L}{\displaystyle\partial v^i_A     } dq^i, \, \, 1\leq A\leq k.
\end{equation}

Introducing the following $2$--forms $(\omega_L)_A =
-d(\theta_L)_A \,, \,  1\leq A \leq k$,
 one can easily prove the following proposition:

\begin{proposition}
 %\label{pr811}
\noindent  $L:T^1_kQ\longrightarrow\r$ is a regular Lagrangian if
and only if $\, ((\omega_L)_1, \dots,(\omega_L)_k)$ is a
polysymplectic structure on $T^1_kQ$.
 \end{proposition}

This polysymplectic structure, associated to $L$, was also
introduced by G\"unther \cite{gun} using the Legendre
transformation.

     The  Legendre map  $ FL: T^1_kQ \longrightarrow (T^1_k)^*Q
$,  was introduced by G\"{u}nther \cite{gun}, and
 we rewrite  it  as follows: if $({v_1}_q, \dots ,
{v_k}_q) \in (T^1_k)_qQ$
$$
[FL({v_1}_q,\ldots , {v_k}_q)]^A(w_q)=
\displaystyle\frac{d}{ds}\displaystyle L({v_{1}}_q, \dots
,{v_{A}}_q+sw_q, \ldots , {v_{k}}_q)_{|s=0}  ,
$$
 for each $1 \leq A \leq k$. We deduce that  $FL$ is locally
given by
\begin{equation}\label{locfl}
(q^i,v^i_A)  \longrightarrow  \left( q^i,
\frac{\displaystyle\partial L}{\displaystyle\partial v^i_A     }
\right).
\end{equation}

  In fact,   from (\ref{betaloc}) and (\ref{locfl}), we easily
obtain  the following Lemma.

 \begin{lemma}\label{le811}
For every $1\leq A\leq k$,  $(\omega_L)_A= (FL)^*(\omega_0)_A$,
where $(\omega_0)_1,\dots,(\omega_0)_k$ are the $2$-forms of the
canonical polysym\-plectic structure or canonical $k$-symplectic
structure of $(T^1_k)^*Q$.
\end{lemma}

Then, from (\ref{locfl}) we get:

 \begin{proposition} Let $L$ be a Lagrangian.
The following conditions are equivalent:

1) $L$ is regular. 2) {\cal FL} is a local diffeomorphism. 3)
$((\omega_L)_1, \dots,(\omega_L)_k)$ is a polysymplectic structure
on $T^1_kQ$.
\end{proposition}

\begin{remark}
{\rm If $FL$ is a global diffeomorphism, then $L$ is called a {\sl
hyper-regular Lagrangian}.}
\end{remark}

\section{$k$-vector fields. Second order partial differential equations on $T^1_kQ$}

\subsection{$k$-vector fields}

Let $M$ be an arbitrary manifold and $\tau_M : T^{1}_{k}M
\longrightarrow M$ its  tangent bundle of $k^1$-velocities.

\begin{definition}  \label{kvector}
A section ${\bf X} : M \longrightarrow T^1_kM$ of the projection
$\tau_M$ will be called a {\rm $k$-vector field} on $M$.
\end{definition}

Since $T^{1}_{k}M$ is  the Whitney sum $TM\oplus
\stackrel{k}{\dots} \oplus TM$ of $k$ copies of $TM$,
  we deduce that a $k$-vector field ${\bf X}$ defines
a family of $k$ vector fields $\{X_{1}, \dots, X_{k}\}$ on $M$ by
projecting  ${\bf X}$ onto every factor. For this reason we will
denote a $k$-vector field ${\bf X}$ by $(X_1, \ldots, X_k)$.

\begin{definition} \label{integsect}
An {\rm integral section}  of the $k$-vector field  \, ${\bf
X}=(X_{1}, \dots, X_{k})$ \, passing through a point $x\in M$ is a
map
  $\phi:U_0\subset \r^k \rightarrow M$, defined on some neighborhood  $U_0$ of $0\in \rk$,
  such that
$$
\phi(0)=x \quad , \quad \phi_*(t)\left(
\displaystyle\frac{\displaystyle\partial}{\displaystyle\partial
t^A}\Big\vert_t \right) = X_{A}(\phi (t)) \, \quad \mbox{for
every} \quad t\in U_0, \quad 1\leq A \leq k,
$$
 or equivalently,    $\phi$ satisfies
$ {\bf X} \circ\phi=\phi^{(1)}, $ where  $\phi^{(1)}$ is the first
prolongation of $\phi$ defined by
$$
\begin{array}{rccl}\label{1prolong}
\phi^{(1)}: & U_0\subset \r^k & \longrightarrow & T^1_kM \\
\medskip
 & t & \longrightarrow & \phi^{(1)}(t)=j^1_0\phi_t \quad ,
 \quad \phi_t ({\bar t})=\phi ({\bar t}+t) \end{array},
$$
for every ${\bar t},t \in \r^k$ such that ${\bar t}+t\in U_0$.
\end{definition}

In local coordinates:
\begin{equation}\label{localfi11}
\phi^{(1)}(t^1, \dots, t^k)=\left( \phi^i (t^1, \dots, t^k),
\displaystyle\frac{\displaystyle\partial\phi^i}{\displaystyle\partial
t^A} (t^1, \dots, t^k)\right), \qquad  1\leq A\leq k\, ,\,  1\leq
i\leq n \, .
\end{equation}

We say that a $k$-vector field ${\bf X}=(X_1,\ldots , X_k)$ on $M$
is integrable if there is an integral section passing through each
point of $M$.

We remark that a $k$-vector field ${\bf X}$ is integrable if, and
only if, $\{X_1,\ldots , X_k \}$ define an involutive distribution
on $M$.

\subsection{Second-order partial differential equations in $T^1_kQ$}

The aim of this subsection is to characterize the integrable
$k$-vector fields on $T^1_kQ$ such that their integral sections
are canonical prolongations of maps from $\r^k$ to $Q$.

In general, if $F:M \to N$ is a differentiable map, then the induced
map $T^1_k(F):T^1_kM \to  T^1_kN$ defined by
$T^1_k(F)(j^1_0g)=j^1_0(F \circ g)$ is given by
$$  T^1_k(F)({v_1}_q,\ldots , {v_k}_q)=(F_*(q){v_1}_q,\ldots
,F_*(q){v_k}_q) \quad ,$$ where ${v_1}_q,\ldots , {v_k}_q\in
T_qQ$, $q\in Q$ , and $F_*(q):T_qM \to T_{F(q)}N$.

\begin{definition} \label{sode0}
A $k$-vector field on $T^1_kQ$, that is, a section ${\bf X}\colon
T^1_kQ\rightarrow T^1_k(T^1_kQ)$ of the projection
$\tau_{T^1_kQ}:T^1_k(T^1_kQ)\rightarrow T^1_kQ$, is  a {\rm second
order partial differential equation ({\sc sopde})} if it is also a
section of the vector bundle
$T^1_k(\tau_Q):T^1_k(T^1_kQ)\rightarrow T^1_kQ$; that is,
\begin{equation}
T^1_k(\tau_Q)\circ{\bf X}=Id_{T^1_kQ} \label{sodedef}
\end{equation}
where $T^1_k(\tau )$ is defined by $T^1_k(\tau_Q
)(j^1_0\gamma)=j^1_0(\tau_Q\circ\gamma)$.
\end{definition}

Let $(q^i)$ be a coordinate system on $Q$ and $(q^i,v^i_A)$ the
induced coordinate system on $T^1_kQ$.  From a direct computation
in local coordinates we obtain that the local expression of a
{\sc sopde} $(X_1 ,\ldots,X_k) $ is
\begin{equation}\label{localsode1}
X_A(q^i,v^i_A)= v^i_A\frac{\displaystyle
\partial} {\displaystyle
\partial q^i}+
(X_A)^i_B \frac{\displaystyle\partial} {\displaystyle
\partial v^i_B},\quad 1\leq A \leq k \quad .
\end{equation}
 If $\varphi:\rk \to
  T^1_kQ$,   is an integral section of $(X_1
,\ldots,X_k) $ locally given by
$\varphi(t)=(\varphi^i(t),\varphi^i_B(t))$ then
$X_A(\varphi(t))=\varphi_*(t)[\partial /\partial t^A(t)]$ and thus
$$
\frac{\displaystyle\partial\varphi^i} {\displaystyle\partial
t^A}(t)=v^i_A(\varphi(t))=\varphi^i_A(t)\, ,\qquad
\frac{\displaystyle\partial\varphi^i_B} {\displaystyle\partial
t^A}(t)=(X_A)^i_B(\varphi(t))\, .
$$

 From (\ref{localfi11}) we obtain the following:

\begin{proposition} \label{sope1}
Let ${\bf X}=(X_1 ,\ldots,X_k) $ be an integrable {\sc sopde}. If
$\varphi$ is an integral section then  $\varphi=\phi^{(1)}$ where
$\phi^{(1)}$ is the first prolongation of the map
$\phi=\tau\circ\varphi:\rk\stackrel{\varphi}{\to}T^1_kQ\stackrel{\tau}{\to}Q$,
and satisfies
\begin{equation}\label{nn1}
 \frac{\displaystyle\partial \phi^i} {\displaystyle\partial t^A
\partial t^B       }(t)= (X_A)^i_B(\phi^{(1)}(t)) \, .
\end{equation}
Conversely, if $\phi:\rk \to Q$ is any map satisfying  (\ref{nn1})
then $\phi^{(1)}$ is an integral section of $(X_1 ,\ldots,X_k) $.
\end{proposition}

\begin{definition}
Let $(X_1 ,\ldots,X_k) $ be an integrable {\sc sopde}.
 A map $\phi:\rk \to Q$ is said to be
a {\rm solution to the {\sc sopde}} if the first prolongation
$\phi^{(1)}$ is an integral section of $(X_1 ,\ldots,X_k) $.

A $k$-vector field which is an integrable {\sc sopde} is called a {\rm
holonomic $k$-vector field}, and its integral sections
$\varphi=\phi^{(1)}$ are called {\rm holonomic sections}.
\end{definition}

Now we show how to characterize the {\sc sopde}'s using the canonical
$k$-tangent structure of $T^1_kQ$.

\begin{definition}
 The  {\rm Liouville vector field} $C$ on $T^1_kQ$ is the infinitesimal generator of the following flow
$$
\begin{array}{ccl}
\r \times T^1_kQ & \longrightarrow &
 T^1_kQ  \\
\noalign{\medskip} (s,({v_1}_q,\ldots , {v_k}_q)) &
\longrightarrow & (e^s \, {v_1}_q,\ldots,e^s \, {v_k}_q)\, ,
\end{array}
$$
and in local coordinates has the form
\begin{equation}\label{localc}
 C =     \sum_{i,B} v^i_B
\displaystyle\frac{\displaystyle\partial}{\displaystyle \partial
v_B^i}.
\end{equation}
\end{definition}

We can write $C=C_1+\ldots+C_k$ where $C_A$, $1\leq A \leq k$, are
the canonical vector fields on $T^1_kQ$ given by the following
flows
$$
\begin{array}{ccl}
\r \times T^1_kQ & \longrightarrow &
 T^1_kQ  \\
\noalign{\medskip} (s,({v_1}_q,\ldots , {v_k}_q)) &
\longrightarrow & ({v_1}_q,\ldots,{v_{A-1}}_q, e^s \,
{v_A}_q,{v_{A+1}}_q, \ldots,{v_k}_q)\, .
\end{array}
$$
In local coordinates
\begin{equation}\label{localca}
 C_A =     \sum_{i} v^i_A
\displaystyle\frac{\displaystyle\partial}{\displaystyle \partial
v_A^i}.
\end{equation}

 From (\ref{localJA}), (\ref{localsode1}), (\ref{localc}) and (\ref{localca}) we deduce the following:

\begin{proposition}
    \label{pr235}
A $k$-vector field ${\bf X}=(X_1,\ldots,X_k)$ on $T^1_kQ$ is a
{\sc sopde}  if, and only if, $S^A(X_A)=C_A$, for all $1\leq A\leq k$,
where $(S^1,\ldots,S^k)$ is the canonical   $k$-tangent structure
on $T^1_kQ$.
\end{proposition}

\section{Hamiltonian and Lagrangian formalism \cite{gun,fam}}

\subsection{Hamiltonian formalism}

Let $(M,\omega_A,V)$ be a $k$--symplectic manifold, and $H:M \to
\r$ a Hamiltonian function. Let ${\bf X}=(X_1,\dots,X_k)$ be a
$k$-vector field on $M$ that satisfies the equations
\begin{equation}\label{geoha}
 \ds\sum_{i=1}^k \,
 \imath_{X_A}\omega_A \,  = \,
dH \, .
\end{equation}

If $X_A$ is locally given by
$$
X_A  =  (X_A)^i \ds\frac{\partial}{\partial
 q^i}
+ (X_A)^i_B\ds\frac{\partial}{\partial v^i_B} \, ,
$$
in a local system of canonical coordinates $(q^i,p^A_i)$, (whose
existence is ensured by the Theorem \ref{thks}) then (\ref{geoha})
is equivalent to the equations
$$
\frac{\displaystyle\partial H}{\displaystyle\partial q^i}=\, - \,
\sum_{A=1}^k\ (X_A)^A_i \quad  , \quad \frac{\displaystyle\partial
H}{\displaystyle\partial p^A_i}=(X_A)^i  \quad .
$$
So if $(X_1,\dots,X_k)$ is also integrable then  its integral
sections $\varphi:\rk \to M$, with
$\varphi(t)=(\varphi^i(t),\varphi^i_A(t))$ are solutions to the
{\it Hamilton-De Donder-Weyl field equations}
\begin{equation}\label{HE}
\frac{\displaystyle\partial H}{\displaystyle\partial q^i}=
-\displaystyle\sum_{A=1}^k\frac{\displaystyle\partial\varphi^A_i}
{\displaystyle\partial t^A}\quad , \quad
\frac{\displaystyle\partial H} {\displaystyle\partial p^A_i}=
\frac{\displaystyle\partial\varphi^i}{\displaystyle\partial t^A},
\quad 1\leq A\leq k, \, 1\leq\ i \leq n\, .
\end{equation}

So, equation  (\ref{geoha}) is a geometric version of the
Hamilton-De Donder-Weyl field equations.

\subsection{Lagrangian  formalism}

In this subsection, we recall the Lagrangian formalism developed
by G\"unther \cite{gun}.

In general, given a Lagrangian function of the form
$L=L(q^i,v^i_A)$, and using a variational principle, one obtains
the {\it Euler-Lagrange field equations} for $L$:
\begin{equation}\label{lageq1}
\displaystyle \sum_{A=1}^k\frac{\displaystyle d}{\displaystyle d
t^A} \left(\frac{\displaystyle\partial L}{\displaystyle \partial
v^i_A}\right)- \frac{\displaystyle \partial L}{\displaystyle
\partial q^i}=0, \qquad v^i_A     =\frac{\displaystyle \partial
 q^i}{\displaystyle \partial t^A}.
\end{equation}

Then, let $L:T^1_kQ\longrightarrow \r$ be a Lagrangian, and let us
consider the $2$-forms $((\omega_L)_1,\dots ,(\omega_L)_k)$ on
$T^1_kQ$ defined by $L$, and $E_L=C(L)-L$, $C$ being the Liouville
vector field in $T^1_kQ$. Now, let ${\bf X}=(X_1,\dots,X_k)$ be a
$k$-vector field in $T^1_kQ$ (that is, a section ${\bf X}\colon
T^1_kQ \longrightarrow T^1_k(T^1_kQ)$) of the projection
$\tau_{T^1_kQ}\colon T^1_k(T^1_kQ)\to T^1_kQ$. Then:

\begin{proposition}    \label{53}
If  ${\bf X}=(X_1,\ldots,X_k)$ is an integrable {\sc sopde}, and
$\psi\equiv\phi^{(1)}\colon\rk \to T^1_kQ$ is an integral section
of ${\bf X}$, then ${\bf X}$ is a solution to the equation
\begin{equation}\label{lageq0}
  \ds\sum_{A=1}^k \, \imath_{X_A}(\omega_L)_A= dE_L \qquad ,
\end{equation}
if, and only if, $\phi\colon\rk \to Q$ is a solution to the
Euler-Lagrange equations (\ref{lageq1}).
\end{proposition}
\proof \quad If each $X_A$ is locally given by
$$
X_A  =  (X_A)^i \ds\frac{\partial}{\partial
 q^i}
+ (X_A)^i_B\ds\frac{\partial}{\partial v^i_B}
 $$
then, from (\ref{betaloc}), (\ref{localc})  and (\ref{lageq0}) we
deduce that $(X_1, \ldots , X_k)$ is a solution to (\ref{lageq0})
if, and only if, $(X_A)^i$ and $(X_A)^i_B$ satisfy the system of
equations
 \begin{equation}\label{locel1}
  \left( \ds\frac{\partial^2 L}{\partial q^i \partial v^j_A} - \ds\frac{\partial^2 L}{\partial q^j \partial v^i_A}
\right) \, (X_A)^j - \ds\frac{\partial^2 L}{\partial v_A^i
\partial v^j_B} \, (X_A)^j_B = v_A^j \ds\frac{\partial^2
L}{\partial q^i
\partial v^j_A} - \ds\frac{\partial  L}{\partial q^i } \, ,
\end{equation}
\begin{equation}\label{locel2}
\ds\frac{\partial^2 L}{\partial v^j_B
\partial v^i_A} \, (X_A)^i=\ds\frac{\partial^2 L}{\partial v^j_B
\partial v^i_A} \, v_A^i \quad .
\end{equation}
But, as ${\bf X}$ is a {\sc sopde}, we have
\begin{equation}\label{locel3}
(X_A)^i= v_A^i,
\end{equation}
then (\ref{locel2}) holds identically, and (\ref{locel1}) is
equivalent to
\begin{equation}\label{locel4}
\ds\frac{\partial^2 L}{\partial q^j \partial v^i_A} v^j_A +
\ds\frac{\partial^2 L}{\partial v_A^i
\partial v^j_B} \, (X_A)^j_B =   \ds\frac{\partial  L}{\partial q^i }
\end{equation}

 Now, if
$\psi(t)=\phi^{(1)}=(\phi^i(t),\phi^i_A(t))$ is an integral
section of ${\bf X}$, then
\begin{eqnarray}
\label{aux11} (X_A)^i(\psi(t))=\phi^i_A(t)=\ds\ds\frac{\partial
\phi^i}{\partial t^A}  \quad ,
\\
\label{aux22} (X_A)^i_B(\psi(t))=\ds\ds\frac{\partial
\phi^i_B}{\partial t^A}=\ds\ds\frac{\partial^2 \phi^i}{\partial
t^A \partial t^B} \quad ,
\end{eqnarray}
and going to (\ref{locel4}) we obtain that
\begin{equation}\label{01}
\ds\frac{\partial^2 L}{\partial q^i \partial v^j_A}(\phi(t))
\ds\frac{\partial\phi^i}{\partial t^A}+\ds\frac{\partial^2
L}{\partial v_B^i
\partial v^j_A}(\phi(t))\ds\frac{\partial\phi^i_B}{\partial t^A}=
\ds\frac{\partial^2 L}{\partial q^i \partial v^j_A}(\phi(t))
\ds\frac{\partial\phi^i}{\partial t^A}+\ds\frac{\partial^2
L}{\partial v_B^i
\partial v^j_A}(\phi(t))\ds\frac{\partial^2 \phi^i}{\partial t^A \partial t^B}
=\ds\frac{\partial  L}{\partial q^i }(\phi(t))
\end{equation}
which are the Euler-Lagrange equations for the map $\phi$.

 Conversely, let ${\bf X}$ be an
integrable {\sc sopde} having
$\psi(t)=\phi^{(1)}=(\phi^i(t),\phi^i_A(t))$ as integral sections,
for every $(\phi^i(t))$ solution to the Euler-Lagrange equations.
Therefore (\ref{aux11}) and (\ref{aux22}) hold since ${\bf X}$ is
a {\sc sopde}, and then (\ref{01}), which holds because
$(\phi^i(t))$ is a solution to the Euler-Lagrange equations, is
equivalent to (\ref{locel4}). Hence ${\bf X}$ is a solution to
(\ref{lageq0}). \begin{flushright}
 \blob \end{flushright}

  In this way, equation (\ref{lageq0}) can be
considered as a geometric version of the Euler-Lagrange field
equations.

Observe that, if the Lagrangian is regular, equation
(\ref{locel2}) leads to conclude that every solution to
(\ref{lageq0}) is a {\sc sopde}. In addition, equation
(\ref{locel4}) leads to defining local solutions to (\ref{lageq0})
in a neighborhood of each point of $T^1_kQ$ and, using a partition
of unity, global solutions to (\ref{lageq0}).

Now let us suppose that the Lagrangian $L:T^1_kQ \to \r$ is
hyper-regular, that is, $FL$ is a diffeomorphism. We consider the
 Hamiltonian $H:$ $(T^1_k)^*Q \to \r$
defined by $H=E_L\circ FL^{-1}$ where $FL^{-1}$ is the inverse map
of $FL$. Then:

\begin{theorem}\label{te421}
a) If ${\bf X}_L=((X_L)_1,\dots,(X_L)_k)$ is a solution to
(\ref{lageq0}) then ${\bf X}_H=((X_H)_1, \ldots, (X_H)_k)$, where
$(X_H)_A=FL_*((X_L)_A)$, $1 \leq A \leq k$, is a solution to
(\ref{geoha}) with $\omega_A=(\omega_0)_A$ and $H=E_L\circ
FL^{-1}$.

b) If ${\bf X}_L=((X_L)_1,\dots,(X_L)_k)$ is integrable,
$\phi^{(1)}$ is an integral section and
$\phi=\tau\circ\phi^{(1)}$, then $\varphi=FL \circ \phi^{(1)}$ is
an integral section of \, ${\bf X}_H=((X_H)_1, \ldots, (X_H)_k)$
and thus it is a solution to the Hamilton-De Donder Weyl equations
(\ref{HE}) for $H=E_L\circ FL^{-1}$ .
\end{theorem}

\proof a) It is an immediate consequence of (\ref{geoha}) and
(\ref{lageq0}) using that $FL^*(\omega_0)_A=(\omega_L)_A$ and
$E_L=H\circ FL^{-1}$.

\noindent b)  It is an immediate consequence of Definition \ref
{integsect} of integral section of a $k$-vector field.
\begin{flushright}
 \blob \end{flushright}

\begin{definition}
A singular Lagrangian system $(T^1_kQ,(\omega_L)_1,\ldots
,(\omega_L)_k)$ is called almost-regular if $\mathcal{P}:= FL
(T^1_kQ)$ is a closed submanifold of $(T^1_k)^*Q$ (we will denote
the natural imbedding by $\jmath_0:\mathcal{P}\hookrightarrow
(T^1_k)^*Q$, $FL$ is a submersion onto its image, and the fibres
$FL^{-1}(FL(v))$, for every $v\in T^1_kQ$,
 are connected submanifolds of $T^1_kQ$.
\end{definition}

In this case there exists $H_0\in\Cinfty({\cal P})$ such that
$(FL_0)^*H_0=E_L$, where $FL_0\colon T^1_kQ\to{\cal P}$ is defined
by $\jmath_0\circ FL_0=FL$, and the Hamiltonian field equation
analogous to (\ref{geoha}) is
\begin{equation}
\ds\sum_{i=1}^k \, \imath_{(X_0)_A}\omega_A^0 = dH_0 \label{HEo}
\end{equation}
where $\omega_A^0=\jmath_0^*(\omega_0)_A$, for every $1 \leq A
\leq k$, and ${\bf X}_0=((X_0)_1,\ldots,(X_0)_k)$ (if it exists)
is a $k$-vector field on ${\cal P}$.

\section{Skinner-Rusk formulation}

\subsection{The Skinner-Rusk formalism for $k$-symplectic field theories}

Let us consider the {\it Whitney sum} $T^1_kQ\oplus_Q (T^1_k)^*Q$,
with coordinates $(q^i,v^i_A,p^A_i)$. It has natural bundle
structures over $T^1_kQ$ and $(T^1_k)^*Q$. Let us denote by
$pr_1:T^1_kQ\oplus_Q (T^1_k)^*Q \to T^1_kQ$ the projection into
the first factor, $pr_1(q^i,v^i_A,p^A_i)=(q^i,v^i_A)$, and
$pr_2:T^1_kQ\oplus_Q (T^1_k)^*Q \to (T^1_k)^*Q$ the projection
into the second factor, $pr_2(q^i,v^i_A,p^A_i)=(q^i,p^A_i)$.

In this bundle, we have some canonical structures. First, let
$((\omega_0)_1, \ldots , (\omega_0)_k)$ be the canonical
polysymplectic structure on $(T^1_k)^*Q$. We shall denote by
$(\Omega_1, \ldots , \Omega_k)$  the pull-back by $pr_2$ of these
$2$-forms to  $T^1_kQ\oplus_Q (T^1_k)^*Q$, that is,
$\Omega_A=(pr_2)^*(\omega_0)_A$, $1 \leq A \leq k$.

Furthermore, the {\it coupling function} in $T^1_kQ\oplus_Q
(T^1_k)^*Q$, denoted by ${\cal C}$, is defined as follows:
$$
\begin{array}{ccccl}
{\cal C} &\colon & T^1_kQ\oplus_Q (T^1_k)^*Q & \longrightarrow & \r \\
& & ({v_1}_q,\ldots ,{v_k}_q,\alpha^1_q,\ldots ,\alpha^k_q)
 & \mapsto & \ds\sum_{A=1}^k \alpha^A_q({v_A}_q)
\end{array}
$$

Given a Lagrangian $L\in\Cinfty(T^1_kQ)$, we can define the {\it
Hamiltonian function} in $T^1_kQ\oplus_Q (T^1_k)^*Q$, denoted by
 $\mathcal{H}\in\Cinfty(T^1_kQ\oplus_Q (T^1_k)^*Q)$, as
$$
\mathcal{H}({v_1}_q,\ldots,{v_k}_q,\alpha^1_q,\ldots,\alpha^k_q)=
{\cal C}({v_1}_q,\ldots,{v_k}_q,\alpha^1_q,\ldots,\alpha^k_q)-
(pr_1^*L)({v_1}_q,\ldots,{v_k}_q,\alpha^1_q,\ldots,\alpha^k_q)
$$
which, in local coordinates, is given by
\begin{equation}\label{s1}
\mathcal{H}= \ds\sum_{A=1}^k \ds\sum_{i=1}^n p^A_i \,
v^i_A-L(q^i,v^i_A) \quad .
\end{equation}

Now, the problem consists in finding the integral sections
$\psi\colon\rk\to T^1_kQ\oplus(T^1_k)^*Q$ of an integrable
$k$-vector field ${\bf Z}=(Z_1,\ldots,Z_k)$ on $T^1_kQ\oplus_Q
(T^1_k)^*Q$, such that
\begin{equation}\label{s3}
\ds\sum_{A=1}^k \, \imath_{Z_A} \Omega_A = d\mathcal{H} \quad .
\end{equation}

Equation (\ref{s3}) gives a different kind of information. In fact,
writing locally each $Z_A$ as
$$
Z_A=  (Z_A)^i \ds\frac{\partial}{\partial q^i} +
(Z_A)^i_B\ds\frac{\partial}{\partial v^i_B}+ (Z_A)^B_i
\ds\frac{\partial}{\partial p^B_i} \, ,
$$
then, from (\ref{canosym}), (\ref{s1}) and (\ref{s3}) we obtain
\begin{equation}\label{s6}
p^A_i= \ds\frac{\partial L}{\partial v^i_A}\circ pr_1
\end{equation}
\begin{equation}
\label{s4} (Z_A)^i=v^i_A
\end{equation}
\begin{equation}\label{s5}
\ds\sum_{A=1}^k (Z_A)^A_i=\ds\frac{\partial L}{\partial q^i}\circ
pr_1
\end{equation}
where $1 \leq A \leq k \, , \, 1 \leq i \leq n$. Then from
(\ref{s4}) we have that $Z_A$ is locally given by
\begin{equation}\label{s7}
Z_A=   v^i_A  \ds\frac{\partial}{\partial q^i} +
(Z_A)^i_B\ds\frac{\partial}{\partial v^i_B}+ (Z_A)^B_i
\ds\frac{\partial}{\partial p^B_i} \quad .
\end{equation}

So, in particular, we have obtained information of three different
classes:
\begin{enumerate}
\item The constraint equations (\ref{s6}), which are algebraic
(not differential) equations defining a submanifold $M_L$ of
$T^1_kQ\oplus_Q(T^1_k)^*Q$ where the equation (\ref{s3}) has
solution. Let us observe that this submanifold is just the graph
of the Legendre map $FL$ defined by the Lagrangian $L$.

We denote by $\jmath\colon M_L \to T^1_kQ\oplus_Q (T^1_k)^*Q$ the
natural imbedding, and by $pr_1^0\colon M_L \to T^1_kQ$ and
$pr_2^0\colon M_L \to (T^1_k)^*Q$ the restricted projections of
$pr_1$ and $pr_2$. \item Equations (\ref{s4}) which are a holonomy
condition similar to (\ref{locel3}) and, as we will see in the
next subsection (see Theorem \ref{ELH-eq}), they force the
integral sections of the $k$-vector field ${\bf Z}$ to be lifting
of sections $\phi:\rk \to Q$. This property is similar to the one
in the unified formalism of Classical Mechanics, and it reflects
the fact that the geometric condition in the unified formalism is
stronger  than the usual one in the Lagrangian formalism.
 \item
Equations (\ref{s5}) which, taking into account (\ref{s6}) and
(\ref{s4}), are just the classical Euler-Lagrange equations (see
Theorem \ref{ELH-eq}).
\end{enumerate}

If ${\bf Z}=(Z_1,\ldots,Z_k)$ is a solution to (\ref{s3}), then
each $Z_A$ is tangent to the submanifold $M_L$ if, and only if,
the functions $Z_A\left(p^B_j -\ds\frac{\partial L}{\partial
v^j_B}\circ pr_1\right)$ vanish at the points of $M_L$, for every
$1 \leq A,B \leq k\, , \, 1 \leq j \leq n$. Then from (\ref{s7})
we deduce that this is equivalent to the following equations
\begin{equation}\label{s8}
 (Z_A)^B_j= v^i_A
\, \ds\frac{\partial^2 L}{\partial q^i\partial v^j_B} + (Z_A)^i_C
\, \ds\frac{\partial^2 L}{\partial v^i_C\partial v^j_B} \quad .
\end{equation}

Thus the problem to be solved is the following:

\begin{state}
To find an integral section $\psi\colon\rk\to M_L\subset
T^1_kQ\oplus(T^1_k)^*Q$ of an integrable $k$-vector field ${\bf
Z}=(Z_1,\ldots,Z_k)$ on $T^1_kQ\oplus_Q (T^1_k)^*Q$ solution to
(\ref{s3}) taking values on $M_L$. (This means that ${\bf Z}$ is
tangent to $M_L$).
\end{state}

\begin{remark} {\rm
\begin{enumerate}
\item Equations (\ref{s3}) have not, in general,  a unique
solution. The solutions to (\ref{s3}) are given by
$(Z_1,\ldots,Z_k)+\ker\Omega^{\sharp}$, where $(Z_1,\dots,Z_k)$ is
a particular solution, and $\Omega^{\sharp}\colon
T^1_k(T^1_kQ\oplus_Q(T^1_k)^*Q) \to T^*(T^1_kQ\oplus_Q
(T^1_k)^*Q)$ is defined as
$\Omega^{\sharp}(Y_1,\ldots,Y_k)=\ds\sum_{A=1}^k
\imath_{Y_A}\Omega_A$. \item If $L$ is regular, then taking into
account (\ref{s4}) and (\ref{s5}) we can define a local $k$-vector
field $(Z_1 ,\ldots , Z_k)$ on a neighborhood of each point in
$M_L$ which is a solution to (\ref{s3}). Each $Z_A$ is locally
given by
$$
(Z_A)^i=v^i_A\quad ,\quad
(Z_A)^B_i=\ds\frac{1}{k}\ds\frac{\partial L}{\partial
q^i}\delta_A^B \, ,
$$
with $(Z_A)^i_B$ satisfying (\ref{s8}). Now, by using a partition
of the unity, one can construct a global $k$-vector field which is
a solution to (\ref{s3}).
\end{enumerate}
}
\end{remark}

 When the Lagrangian function $L$ is singular we cannot assure
the existence of consistent solutions for equation (\ref{s3}).
Then we must develop a constraint algorithm for obtaining a
constraint submanifold (if it exists) where these solutions exist.
Next, we outline this procedure (see also \cite{LMM-2002}, where a
similar algorithm is sketched in the multisymplectic formulation).

First, in order to assure the existence of a Hamiltonian
counterpart for the singular Lagrangian system we
assume, from now on, that the singular Lagrangians are almost-regular.

We begin with $P_0=M_L$. Then, let $P_1$ be the subset of $P_0$
made of those points where there exists a solution to (\ref{s3}),
that is,
$$
P_1= \{ z \in P_0 \, |\,  \exists (Z_1, \ldots, Z_k) \in (T^1_k)_z
P_0 \mbox{ solution to } (\ref{s3}) \}
$$
If $P_1$ is a submanifold of $P_0$, then there exists a section of
the canonical projection $\tau_{P_0}:T^1_k P_0 \to P_0$ defined on
$P_1$ which is a solution to (\ref{s3}), but that does not define,
in general, a $k$-vector field on $P_1$. To find solutions taking
values into $T^1_k P_1$, we define a new subset $P_2$ of $P_1$ as
follows
$$
P_2= \{ z \in P_1 \, |\,  \exists (Z_1, \ldots, Z_k) \in (T^1_k)_z
P_1 \mbox{ solution to } (\ref{s3}) \}
$$
If $P_2$ is a submanifold of $P_1$, then there exists a section of
the canonical projection $\tau_{P_1}:T^1_k P_1 \to P_1$ defined on
$P_2$ which is a solution to (\ref{s3}), but that does not define,
 in general, a $k$-vector field on $P_2$.

Procceding further, we get a family of constraint manifolds
$$
\ldots \, \hookrightarrow \, P_2 \, \hookrightarrow \, P_1 \,
\hookrightarrow \,  P_0=M_L \, \hookrightarrow \, (T^1_k)^*Q\oplus
T^1_kQ
$$
If there exists a natural number $f$ such that $P_{f+1}=P_f$ and
$dim \, P_f >k$ then we call $P_f$ the {\it final constraint
submanifold} over which we can find solutions to equation
(\ref{s3}). Let us observe that the solutions will not be unique
(even in the regular case) and, in general, will not be
integrable.
 In order to find integrable solutions to equation (\ref{s3}),
a constraint algorithm based on the same idea must be developed.

\subsection{The field equations for sections}

$M_L$ being the graph of $FL$, it is diffeomorphic to $T^1_kQ$ (so
$pr_1^0$ is a diffeomophism). Let ${\bf Z}=(Z_1, \ldots , Z_k)$ be
an integrable $k$-vector field solution to (\ref{s3}). Every
integral section $\psi\colon t\in\rk \to
(\psi^i(t),\psi^i_A(t),\psi^A_i(t))\in T^1_kQ\oplus_Q (T^1_k)^*Q$
of ${\bf Z}$ solution to (\ref{s3}) is of the form
$\psi=(\psi_L,\psi_H)$, with $\psi_L=pr_1\circ\psi\colon\rk\to
T^1_kQ$, and if $\psi$ takes values in $M_L$ then
$\psi_H=FL\circ\psi_L$ ; in fact, from (\ref{s6}) we obtain
$$
\psi_H(t)=(pr_2 \circ \psi)(t)=(\psi^i(t),\psi^A_i(t))
=\left(\psi^i(t), \ds\frac{\partial L}{\partial
v^i_A}\Big\vert_{\psi_L(t)}\right) =(FL \circ \psi_L)(t) \, .
$$

In this way, every constraint, differential equation, etc. in the
unified formalism can be translated to the Lagrangian or the
Hamiltonian formalisms by restriction to the first or the second
factors of the product bundle. In particular, conditions
(\ref{s6}) generate, by $pr_2$-projection, the primary constraints
of the Hamiltonian formalism for singular Lagrangians (i.e., the
image of the Legendre transformation,
$FL(T^1_kQ)\subset(T^1_k)^*Q$ , and they can be called {\it
primary Hamiltonian constraints}.

In this way the main result in this subsection is the following:

\begin{theorem}\label{ELH-eq}
Let ${\bf Z}=(Z_1,\ldots, Z_k)$ be an integrable $k$-vector field
in $T^1_kQ\oplus_Q (T^1_k)^*Q$ solution to (\ref{s3}) and let
$\psi\colon\rk\to M_L\subset T^1_kQ\oplus_Q(T^1_k)^*Q$ be an
integral section of ${\bf Z}=(Z_1,\ldots,Z_k)$, with
$\psi=(\psi_L,\psi_H)=(\psi_L,FL\circ\psi_L)$. Then $\psi_L$ is
the canonical lift $\phi^{(1)}$ of the projected section
$\phi=\tau_Q\circ pr_1\circ\psi\colon\rk \to Q$,
 and $\phi$ is a solution to the Euler-Lagrange field
equations (\ref{lageq1}).
\end{theorem}

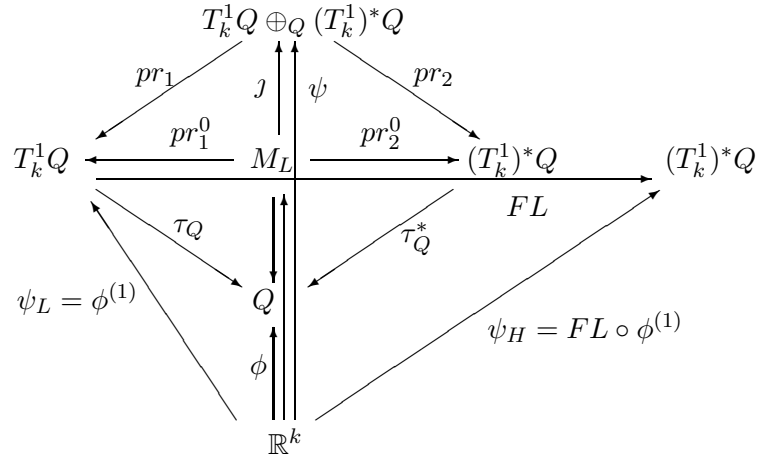
$$
\begin{array}{cccc}
& \begin{picture}(135,50)(0,0) \put(40,30){\mbox{$T^1_kQ\oplus_Q
(T^1_k)^*Q$}} \put(15,10){\mbox{$pr_1$}}
\put(120,10){\mbox{$pr_2$}} \put(55,25){\vector(-3,-2){55}}
\put(90,25){\vector(3,-2){55}} \put(60,6){\mbox{$\jmath$}}
\put(69,-10){\vector(0,1){35}}
\end{picture} & &
\\
T^1_kQ &
\begin{picture}(135,20)(0,0)
\put(28,10){\mbox{$pr^0_1$}} \put(52,3){\vector(-1,0){56}}
\put(58,0){\mbox{$M_L$}} \put(100,10){\mbox{$pr^0_2$}}
\put(81,3){\vector(1,0){56}}
\end{picture}
&
\begin{picture}(80,20)(0,0)
\put(-5,0){\mbox{$(T^1_k)^*Q$}} \put(70,0){\mbox{$(T^1_k)^*Q$}}
\end{picture}
\\ &
\begin{picture}(135,105)(0,0)
 \put(0,104){\vector(1,0){210}}
\put(155,90){\mbox{$FL$}} \put(29,84){\mbox{$\tau_Q$}}
\put(115,80){\mbox{$\tau^*_Q$}}
\put(-30,55){\mbox{$\psi_L=\phi^{(1)}$}}
 \put(148,44){\mbox{$\psi_H=FL\circ\phi^{(1)}$}}
 %\put(77,32){\mbox{$\psi_0$}}
\put(80,135){\mbox{$\psi$}}
 \put(58,30){\mbox{$\phi$}}
\put(59,55){\mbox{$Q$}}
 \put(65,0){\mbox{$\rk$}}
\put(67,97){\vector(0,-1){32}} \put(0,100){\vector(3,-2){55}}
 \put(135,100){\vector(-3,-2){55}}
\put(53,13){\vector(-2,3){55}}
 \put(83,13){\vector(3,2){130}}
\put(67,13){\vector(0,1){35}}
 \put(71,13){\vector(0,1){85}}
 \put(75,13){\vector(0,1){143}}
\end{picture} &
\end{array}
$$
\proof If $\psi(t)=(\psi^i(t),\psi^i_A(t),\psi^A_i(t))$ is an
integral section of ${\bf Z}=(Z_1,\ldots, Z_k)$, then
\begin{equation}
Z_A(\psi(t))=\ds\frac{\partial\psi^i }{\partial t^A}(t)
\ds\frac{\partial}{\partial q^i}\Big\vert_{\psi(t)} +
\ds\frac{\partial \psi^B_i}{\partial t^A}(t)
\ds\frac{\partial}{\partial p^B_i}\Big\vert_{\psi(t)} +
\ds\frac{\partial\psi^i_B}{\partial t^A}(t) \ds\frac{\partial
}{\partial v^i_B}\Big\vert_{\psi(t)} \label{aux0}
\end{equation}
 From  (\ref{s6}), (\ref{s4})   and (\ref{aux0}) we obtain
\begin{eqnarray}
\psi^i_A(t) & = & v^i_A(\psi(t)) =
(Z_A)^i(\psi(t))=\ds\frac{\partial \psi^i }{\partial t^A}(t)
 \label{aux1}
\\
\psi^A_i(t)& = & p^A_i(\psi(t))= \left(\frac{\partial L}{\partial
v^i_A}\circ pr_1\right)(\psi(t))= \frac{\partial L}{\partial
v^i_A}\Big\vert_{\psi_L(t)} \label{aux2}
\\
\ds\frac{\partial \psi^B_i }{\partial t^A}(t) & = &
(Z_A)^B_i(\psi(t)) , \label{aux4}
\end{eqnarray}

Therefore from (\ref{s5}), (\ref{aux2})  and (\ref{aux4})  we
obtain
$$
\ds\frac{\partial L}{\partial q^i}(\psi_L(t))   = \ds\sum_{A=1}^k
(Z_A)^A_i(\psi(t))   = \ds\sum_{A=1}^k\ds\frac{\partial
\psi^A_i}{\partial t^A}(t)= \ds\sum_{A=1}^k\ds\frac{\partial}
{\partial t^A} \left(\ds\frac{\partial L}{\partial
v^i_A}\Big\vert_{ \psi_L(t)}\right)
$$
and from (\ref{aux1})
$$
\psi^i_A(t) = \ds\frac{\partial \psi^i }{\partial t^A} (t) \, .
$$

The last two equations are the Euler-Lagrange field equations for
the section $\phi(t)=(\psi^i(t))=(\tau \circ pr_1 \circ\psi)(t)$,
and $\psi_L=\phi^{(1)}$. \begin{flushright}
 \blob \end{flushright}

In addition, for the regular case we can prove:

\begin{proposition}
Under the hypothesis of Theorem \ref{ELH-eq}, if $L$ is regular
then $\psi_H=FL\circ\psi_L$ is a solution to the Hamilton-De
Donder-Weyl field equations (\ref{HE}), where the Hamiltonian $H$
is locally given by $H \circ FL = E_L$. \label{ELH-eqbis}
\end{proposition}
\proof Since $L$ is regular, $FL$ is a local diffeomorphism  and
thus we can choose for each point in $T^1_kQ$ an open neighborhood
$U \subset T^1_k Q$ such that $FL_{\vert U}:U \to FL(U)$ is a
diffeomorphism. So we can define $H_U:FL(U) \to \r$ as $H_U=
(E_L)_{\vert U} \circ (FL_{\vert U})^{-1}$.

Denoting by $H\equiv H_U$, $E_L\equiv(E_L)_{\vert U}$ and
$FL\equiv FL_{\vert U}$, we have $E_L=H\circ FL$   which we
provides the identities
\begin{equation}\label{sh0}
\ds\frac{\partial H}{\partial p^A_i}\circ FL =v^i_A \quad , \quad
\ds\frac{\partial H}{\partial q^i} \circ FL= \, - \,
\ds\frac{\partial L}{\partial q^i} \quad .
\end{equation}

Now considering the open subset $V=\psi_L^{-1} (U) \subset \r^k$
  we have $\psi_{\vert V} : V \subset \r^k \to U \oplus
FL(U) \subset M_L$, where $(\psi_L)_{\vert V} : V \subset \r^k \to
U \subset T^1_k Q$ and $(\psi_H)_{\vert V} = FL \circ
(\psi_L)_{\vert V} : V \subset \r^k \to FL(U) \subset (T^1_k)^*
Q$.

Therefore from (\ref{s5}), (\ref{aux1}), (\ref{aux4}) and
(\ref{sh0}), for every $t \in V \subset \r^k$ we obtain
$$
\ds\frac{\partial H}{\partial p^A_i}\Big\vert_{\ds\psi_H(t)}=
\left(\ds\frac{\partial H}{\partial p^A_i}\circ
FL\right)(\psi_L(t))= v^i_A(\psi_L(t)) =
\ds\frac{\partial\psi^i}{\partial t^A}(t)
$$
and
$$
\ds\frac{\partial H}{\partial q^i}\Big\vert_{\ds\psi_H(t)}= \left(
\ds\frac{\partial L}{\partial q^i} \circ FL \right)(\psi_L(t))= -
\ds\frac{\partial L}{\partial q^i}\Big\vert_{\psi_L(t)} =
-\sum_{A=1}^k (Z_A)^A_i(\psi(t))= -\sum_{A=1}^k \ds\frac{\partial
\psi^A_i}{\partial t^A}(t)
$$
from which we deduce that $(\psi_H)_{\vert V}$ is a solution to
the Hamilton-De Donder-Weyl field equations
(\ref{HE}).\begin{flushright}
 \blob \end{flushright}

Conversely, we can state:

\begin{proposition}\label{33}
If $L$ is regular and ${\bf X}=(X_1, \ldots, X_k)$ is a  solution
to (\ref{lageq0}) then:
\begin{enumerate} \item The $k$-vector field
${\bf Z}=(Z_1, \ldots, Z_k)$ given by $Z_A=(Id_{T^1_kQ} \oplus
FL)_*(X_A)\, , \, 1 \leq A \leq k$ is a solution to (\ref{s3}).

\item If $\psi_L:  \rk \to   T^1_kQ$ is an integral section of
${\bf X}=(X_1, \ldots, X_k)$ (and thus, from Proposition \ref{53},
a solution to the Euler-Lagrange field equations) then
$\psi=(\psi_L, FL \circ \psi_L): \rk \to   M_L \subset T^1_kQ
\oplus_Q (T^1_k)^*Q $ is an integral section of ${\bf Z}=(Z_1,
\ldots, Z_k)$.
\end{enumerate}
\end{proposition}

\proof  \begin{enumerate} \item If $L$ is regular and ${\bf
X}=(X_1, \ldots, X_k)$ is a  solution to (\ref{lageq0}), then from
Proposition \ref{53} we know that $X_A$ is a {\sc sopde} and thus $X_A$
is locally given by
\begin{equation}\label{xi11}
X_A = v^i_A \, \ds\frac{\partial}{\partial q^i}+ (X_A)^i_B
\ds\frac{\partial}{\partial v^i_B}
\end{equation}
where $(X_A)^i_B$ satisfy (\ref{locel4}). Since the map
$Id_{T^1_kQ} \oplus FL: T^1_kQ \to M_L \subset T^1_kQ \oplus
(T^1_k)^*Q$, is locally given by
\begin{equation}\label{locmapx}(q^i,v^i_A) \to
\left( q^i,v^i_A, \ds\frac{\partial L}{\partial v^i_A} \right),
\end{equation}
from (\ref{xi11}) and (\ref{locmapx}) we obtain
\begin{equation}\label{idea}
Z_A=(Id_{T^1_kQ} \oplus FL)_*(X_A)= v^i_A
\ds\frac{\partial}{\partial q^i} + \left(v^i_A \ds\frac{\partial^2
L }{\partial q^i \partial v^j_C}+(X_A)^i_B \ds\frac{\partial^2 L
}{\partial v^i_B
\partial v^j_C} \right) \ds\frac{\partial}{\partial p^C_j}+
(X_A)^i_B\ds\frac{\partial}{\partial v^i_B}
\end{equation}

Then from (\ref{locel4}) and (\ref{idea}) we have that
$$
\ds\sum_{A=1}^k (Z_A)^A_j=  v^i_A \ds\frac{\partial^2 L }{\partial
q^i \partial v^j_A}+(X_A)^i_B  \ds\frac{\partial^2 L }{\partial
v^i_B
\partial v^j_A}= \ds\frac{\partial L}{\partial
q^j}\, , \quad (Z_A)^i=v^i_A \, , \quad Z_A \left(p^B_k -
\ds\frac{\partial L}{\partial v^k_B}\right)=0 \, ,
$$
that is, the $k$-vector field ${\bf Z}=(Z_1,\ldots, Z_k)$ is
a solution to (\ref{s3}) and each $Z_A$ is tangent to $M_L$  for
$A:1 , \ldots ,  k$.

\item  It follows from Definition \ref{integsect} taking into
account that $pr_2 \circ \psi \, = \, FL \circ \psi_L$.
\end{enumerate}
\begin{flushright}
 \blob \end{flushright}

\begin{remark}
{\rm The last result really holds for regular and almost-regular
Lagrangians. In the almost-regular case, the proof is the same,
but the sections $\psi$, $\psi_L$ and $\psi_H$ take values not on
$M_L$, $T^1_kQ$ and $(T^1_k)^*Q$, but in the final constraint
submanifold $P_f$ and on the projection submanifolds
$pr_1(P_f)\hookrightarrow T^1_kQ$ and $pr_2(P_f)\hookrightarrow
(T^1_k)^*Q$, respectively.}
\end{remark}

\subsection{The field equations for $k$-vector fields}

The aim of this subsection is to establish the relationship between
$k$-vector fields that are solutions to (\ref{lageq0}) and $k$-vector fields
that are solutions to (\ref{s3}). The main result is the following:

\begin{theorem}\label{ELH-kvf}
Let ${\bf Z}=(Z_1,\ldots, Z_k)$ be a $k$-vector field on $M_L$
solution to (\ref{s3}). Then the $k$-vector field ${\bf
X}_L=((X_L)_1,\ldots, (X_L)_k)$ on $T^1_k Q$ defined by
\begin{equation} \label{xlz}
{\bf X}_L\circ pr^0_1=T^1_k (pr^0_1)\circ{\bf Z}
\end{equation}
is a $k$-vector field solution to (\ref{lageq0}) (where $T^1_k
(pr^0_1)\colon T^1_k(M_L) \to T^1_k(T^1_kQ)$ is the natural
extension of $(pr^0_1)_*$).

Conversely, every $k$-vector field ${\bf X}_L$ solution to
(\ref{lageq0}) can be recovered in this way from a $k$-vector
field ${\bf Z}$ in $M_L$ solution to (\ref{s3}).

Moreover, the $k$-vector field ${\bf Z}$ is integrable iff the
$k$-vector field ${\bf X}_L$ is holonomic.
\end{theorem}

\proof Since $pr^0_1:M_L \to T^1_kQ$ is a
diffeomorphism, then the $k$-vector field ${\bf X}_L$ on $T^1_kQ$
defined by (\ref{xlz}) is given by
\begin{equation}\label{xaz}
(X_L)_A \, = \, \left( (pr^0_1)^{-1} \right)^* \, Z_A \, , \quad 1
\leq A \leq k \, .
\end{equation}

Now, for every $1 \leq A \leq k$ we have that
\begin{equation}\label{jomega}
 \jmath^*\Omega_A = (pr^0_1)^* (\omega_L)_A \quad ,
\end{equation}
which follows from  Lemma 2.1 $$\begin{array}{lcl}
\jmath^*\Omega_A & = & \jmath^* (pr_2)^* (\omega_0)_A \, = \,
(pr^0_2)^* (\omega_0)_A \, = \, (FL \circ pr^0_1)^* (\omega_0)_A
\\ \noalign{\medskip}
  & = & (pr^0_1)^* FL^* (\omega_0)_A \, = \, (pr^0_1)^*
  (\omega_L)_A \, .
\end{array}$$

On the other hand  we obtain that
\begin{equation}\label{jesth}
\jmath^*\mathcal{H}=(pr^0_1)^* E_L \quad ,
\end{equation}
from the following computation
$$
\begin{array}{lcl}
\jmath^*\mathcal{H} & = & \jmath^* ( \mathcal{C} - (pr_1)^*L ) \,
= \, \jmath^*\mathcal{C} \, - \, \jmath^* (pr_1)^*L
\\ \noalign{\medskip}
  & = & (pr^0_1)^* CL \, - \, (pr^0_1)^*L \, = \, (pr^0_1)^*E_L \, .
\end{array}$$
 From (\ref{xaz}) and (\ref{jomega}) we deduce that
\begin{equation}\label{sumas}
\ds\sum_{A=1}^k  \imath_{Z_A} \jmath^* \Omega_A \, = \,
\ds\sum_{A=1}^k \imath_{(pr^0_1)^*(X_L)_A} (pr^0_1)^*(\omega_L)_A
\, = \, (pr^0_1)^* \left( \ds\sum_{A=1}^k  \imath_{(X_L)_A}
(\omega_L)_A \right) \, ,
\end{equation}
and from (\ref{jesth}) we deduce that
\begin{equation}\label{dif}
d \left( \jmath^* \mathcal{H} \right) \, = \, d \left( (pr^0_1)^*
E_L \right) \, = \, (pr^0_1)^* dE_L \, .
\end{equation}

Since $pr^0_1$ is a diffeomorphism, from (\ref{sumas}) and
(\ref{dif}) we deduce that the $k$-vector field ${\bf Z}$ is
a solution to (\ref{s3}) iff the $k$-vector field ${\bf X}_L$ is
a solution to (\ref{lageq0}).

Let us suppose now that the $k$-vector field ${\bf Z}$ is
integrable. As a consequence of Theorem \ref{ELH-eq}, for every
integral section $\psi=(\psi_L,FL \circ \psi_L)$ of ${\bf Z}$,
$\psi_L=\phi^{(1)}$, for $\phi=\tau\circ pr_1\circ\psi$. Then
$$
(X_L)_A(pr_1^0(\psi(t)))=(pr_1^0)_*(\psi(t))(Z_A(\psi(t)))=(pr_1^0
\circ \psi)_*(t)\left( \ds\frac{\partial}{\partial
t^A}\Big\vert_q\right)=(\psi_L)_*(t)\left(\ds\frac{\partial}{\partial
t^A}\Big\vert_q\right)
$$

So, $\psi_L=\phi^{(1)}$ is an  integral section of ${\bf X}_L$,
and hence ${\bf X}_L$ is holonomic.

Conversely, if ${\bf X}_L$ is holonomic then for every integral
section $\psi_L=\phi^{(1)}$ with $\phi:\r^k \to Q$, the map
$\psi=(\psi_L,FL \circ \psi_L)$ is an integral section of ${\bf
Z}$. In fact, from (\ref{xaz}), for every $1 \leq A \leq k$
$$
\begin{array}{lcl}
Z_A(\psi(t)) & = & \left( (pr^0_1)^* (X_L)_A \right) (\psi(t)) \,
= \, \left( (pr^0_1)^{-1} \right)_* (\psi_L(t)) \left(
(X_L)_A(\psi_L(t)) \right) \\ \noalign{\medskip}
  & = & \left( (pr^0_1)^{-1} \right)_* (\psi_L(t))
  \left( (\psi_L)_*(t)\left( \ds\frac{\partial}{\partial t^A}(t) \right)
\right) \, = \, \left( (pr^0_1)^{-1} \circ \psi_L \right)_* (t)
\left( \ds\frac{\partial}{\partial t^A}\Big\vert_q \right) \\
\noalign{\medskip}
  & = & \psi_* (t)
\left( \ds\frac{\partial}{\partial t^A}\Big\vert_q \right) \, .
\end{array}
$$
\begin{flushright}
 \blob \end{flushright}

If $L$ is regular, in a neighborhood of each point of $T^1_kQ$
there exists a local solution ${\bf X}_L=((X_L)_1, \ldots ,
(X_L)_k)$ to (\ref{lageq0}). As $L$ is regular, $FL$ is a local
diffeomorphism, so this open neighborhood can be chosen in such a
way that  $FL$ is a diffeomorphism onto its image. Thus
in a neighborhood of each point of $FL(T^1_kQ)$ we can define
$$
(X_H)_A=[(FL)^{-1}]^*(X_L)_A\, ,\quad 1\leq A \leq k\, .
$$
or equivalently, in terms of $k$-vector fields
$$T^1_k(FL) \circ X_L \, = \, X_H \quad .$$
\begin{proposition}\begin{enumerate}
\item The local $k$-vector field ${\bf X}_H=((X_H)_1,\ldots ,(X_H)_k)$
is
  a solution to (\ref{geoha}), where the Hamiltonian $H$ is
locally given by $H \circ FL = E_L$. (In other words, the local
$k$-vector fields ${\bf X}_L$ and ${\bf X}_H$ solution to
(\ref{lageq0}) and (\ref{geoha}), respectively, are $FL$-related).

\item Every local integrable $k$-vector field solution to
(\ref{geoha}) can be recovered in this way from a local integrable
$k$-vector field ${\bf Z}$ in $T^1_kQ\oplus_Q (T^1_k)^*Q$ solution
to (\ref{s3}). \label{ELH-kvfbis}
\end{enumerate}
\end{proposition}

\proof \begin{enumerate} \item This is the local version of
Theorem \ref{te421} a).

\item On the other hand, if ${\bf X}_H$ is a local integrable
$k$-vector field  solution to (\ref{geoha}), then we can obtain the
$FL$-related local integrable $k$-vector field ${\bf X}_L$
solution to (\ref{lageq0}). By Theorem \ref{ELH-kvf}, we recover
${\bf X}_L$ by a local integrable $k$-vector field ${\bf Z}$
solution to (\ref{s3}). \end{enumerate}
\begin{flushright}
 \blob \end{flushright}

\section{Field operators}

\subsection{The evolution operator ${\cal K}$ in mechanics}
\protect\label{pK}

The so-called {\sl time-evolution ${\cal K}$-operator} in
mechanics (also known by some authors as the {\sl relative
Hamiltonian vector field} \cite{PV-00}) is a tool which has mainly
been developed in order to study the Lagrangian and Hamiltonian
formalisms for singular mechanical systems and their equivalence.
It was first introduced in a non-intrinsic way in
\cite{BGPR-eblhf} as an ``evolution operator'' to connect both
formalisms.

In Classical Mechanics, the evolution operator ${\cal K}$
associated with a Lagrangian $L:TQ \to \r$  is a map $\K :TQ \to
T(T^*Q)$ satisfying the following conditions (see \cite{GP-01}):
\begin{enumerate}
\item (Structural condition): $K$  is a vector field along $FL$,
that is, $\tau_{T^*Q} \circ {\cal K} = FL$, where $FL$ is the
Legendre map defined by $L$ and $\tau_{T^*Q}:T(T^*Q) \to T^*Q$ is
the natural projection. \item
 (Dynamical condition): $(FL)^*(\imath _{{\cal K}}(\omega
\circ FL))=dE_L$, where $\omega$ is the canonical symplectic form
on $T^*Q$ and $E_L=CL-L$, being $C$ the Liouville vector field on
$TQ$. \item (Second-order condition): $T(\tau^*) \circ {\cal
K}=Id_{TQ}$, where $\tau^*:T^*Q \to Q$ is the canonical
projection.
\end{enumerate}

 The existence and uniqueness of this operator is studied
in \cite{GP-01}. Its local expression is
$$
{\cal K} = v^i \left( \displaystyle\frac{\partial}{\partial q^i}
\circ FL \right) + \displaystyle\frac{\partial L}{\partial q^i}
\left( \displaystyle\frac{\partial}{\partial p_i}\circ FL
\right)\, .
$$

By definition $\varphi: \r \to TQ$ is an integral curve of ${\cal
K}$ if
\begin{equation}\label{K1}
T(FL) \circ \stackrel{\bullet}{\varphi}= {\cal K} \circ \varphi\,
,
\end{equation}
where $\stackrel{\bullet}{\varphi}:\r \to T(TQ)$ is the
prolongation of $\varphi$ to the tangent bundle $T(TQ)$ of $TQ$.
So we have the diagram \vspace{1cm}
\begin{center}
\begin{picture}(200,80)(0,0)
\put(15,15){\makebox(0,0){$TQ$}} \put(125,15){\makebox(0,0){$
T^*Q$}}
 \put(60,20){$FL$}
\put(30,15){\vector(1,0){80}} \put(125,100){\vector(0,-1){70}}
\put(125,110){\makebox(0,0){$T(T^*Q)$}}
\put(30,25){\vector(1,1){80}} \put(130,60){$\tau_{T^*Q}$} \put(60
,70 ){${\cal K}$} \put(15,110) {\makebox(0,0){$T(TQ)$}}
\put(20,100){\vector(0,-1){70}} \put(40,110 ){\vector(1,0){60}}
\put(25,70){$\tau_{TQ}$}
 \put(60,115){$T(FL)$}
\put(-75,25){\vector(1,1){80}} \put(-60,70
){$\stackrel{\bullet}{\varphi}$}
 \put(-75,15){\makebox(0,0){$\r$}}
 \put(-35,20){\makebox(0,0){$\varphi$}}
 \put(-70,15){\vector(1,0){70}}
\end{picture}
\end{center}

\noindent Moreover, $\varphi=\stackrel{\bullet }{\phi}$, for
$\phi: \r \to Q$, that is, $\varphi$ is holonomic.

The most relevant properties of this operator  are the following:
\begin{itemize}
\item
 If there exists an Euler-Lagrange vector
field $X_L$ on $TQ$, that is, a solution to the equation
$\imath_{X_L}\omega_L=dE_L$, then $\varphi:\r \to TQ$ is an
integral curve of $X_L$ if, and only if, it is an integral curve
of ${\cal K}$; that is, relation (\ref{K1}) holds.

 As a direct consequence of this fact, the relation between ${\cal K}$
 and $X_L$ is
\begin{equation}\label{K2a}
T(FL) \circ X_L={\cal K} \, .
\end{equation}
In general, if the dynamical system is singular, the
Euler-Lagrange vector fields exist only on a submanifold
$S\hookrightarrow TQ$.
\item
 If there exists a Hamilton-Dirac vector field
$X_H$ on $T^*Q$ associated with the the Lagrangian system
$(TQ,\omega_L,E_L)$ (that is, a vector field solution to the
Hamilton-Dirac equations in the Hamiltonian formalism), then
$\psi:\r \to T^*Q$ is an integral curve of $X_H$ if, and only if,
\begin{equation}\label{K3}
\stackrel{\bullet}{\psi} =   K \circ  T(\tau^*_{Q}) \circ
\stackrel{\bullet}{\psi} \, .
\end{equation}
As a consequence, the relation between ${\cal K}$ y $X_H$ is
\begin{equation}\label{K4}
X_H \circ   FL={\cal K}   \, .
\end{equation}
\item
 If $\xi \in C^\infty(T^*Q)$ is a Hamiltonian
constraint, then $\imath_K(d\xi \circ FL)$ is a Lagrangian
constraint.
\end{itemize}

  Relations (\ref{K1}),
(\ref{K2a}),(\ref{K3}) and (\ref{K4}) show how the Lagrangian and
Hamiltonian descriptions can be unified by means of the operator
${\cal K}$.

Some relevant results obtained using this operator are:
\begin{itemize}
\item
The equivalence between the Lagrangian and Hamiltonian
formalisms is proved by means of this operator in the following
way: there is a bijection between the sets of solutions of
Euler-Lagrange equations and Hamilton equations, even though the
dimensions of the final constraint submanifold in both formalisms
are not the same, in general \cite{BGPR-eblhf}, \cite{GP-ggf}.
\item
The complete classification of constraints is achieved. All
the Lagrangian constraints can be obtained from the Hamiltonian
ones using the ${\cal K}$-operator \cite{BGPR-eblhf}.
\item
Noether's theorem is proved and the relation between the
generators of gauge and ``rigid'' symmetries in the Lagrangian and
Hamiltonian formalisms is studied \cite{FP-90}, \cite{GaP-00},
\cite{GP-92b}, \cite{GP-00}.
\item
This operator has been applied
to studying Lagrangian systems whose Legendre map has {\sl generic
singularities}; that is, it degenerates on a hypersurface
\cite{PV-00}, \cite{PV-00b}.
\end{itemize}

\subsection{Field operators $\mathcal{K}$ in field theories}

Next we generalize the definition, properties and some of the
applications of the evolution operator for the $k$-symplectic
formulation of field theories, in order to describe the
relationship between the Lagrangian and Hamiltonian formalisms (
the generalization for the multisymplectic formulation is given
\cite{bar5}). In particular, we will study how to obtain the
solutions of Lagrangian and Hamiltonian field equations by means
of this operator, and the relation between them.

\begin{definition}\label{def61}
A {\rm field operator} $\K$ associated with a Lagrangian $L:T^1_kQ
\to \r$ is a map
$$\K: T^1_kQ \to T^1_k((T^1_k)^*Q)$$ satisfying the
following conditions:
\begin{enumerate}
\item Structural condition : $\K$ is a $k$-vector field along
$FL$, that is
\begin{equation}\label{evo1}
\tau_{(T^1_k)^*Q} \circ \K = FL \, .
\end{equation}
\vspace{1cm}
\begin{center}
\begin{picture}(200,80)(0,0)
\put(15,15){\makebox(0,0){$T^1_kQ$}} \put(180,15){\makebox(0,0){$
(T^1_k)^*Q$}}
 \put(90,25){$FL$}
\put(50,15){\vector(1,0){100}} \put(180,100){\vector(0,-1){70}}
\put(180,110){\makebox(0,0){$T^1_k((T^1_k)^*Q)$}}
\put(25,30){\vector(2,1){140}} \put(190,60){$\tau_{(T^1_k)^*Q}$}
\put(70 ,70 ){$\K$}
\end{picture}
\end{center}
Hence $\K=(\K_1,\ldots ,\K_k)$, where each ${\cal K}_A$, $1 \leq A
\leq k$, is a vector field along $FL$. \item
  Field equation condition:
\begin{equation}\label{evo2}
   \sum_{A=1}^k
(FL)^*[\imath _{\K_A}(\omega_0)_A \circ FL)]=dE_L \, .
\end{equation}
\item
  Second-order condition:
\begin{equation}\label{evo3}
T^1_k(\tau_Q^*) \circ \K =Id_{T^1_kQ} \, .
\end{equation}
\end{enumerate}
\label{tildek}
\end{definition}

Now we are going to calculate the local expression of a field
operator $\K$. If $v=({v_1}_q, \ldots , {v_k}_q) \in T^1_k Q$ then
from (\ref{evo1}) we have that
$$
\K_A(v)= (\K_A)^i(v) \displaystyle\frac{\partial}{\partial
q^i}\Big\vert_{FL(v)}+ (\K_A)^B_i(v)
\displaystyle\frac{\partial}{\partial p^B_i}\Big\vert_{FL(v)}\, ,
\quad 1\leq A \leq k \, .
$$

Taking into account (\ref{evo3}) and that the map
$T^1_k(\tau^*_Q): T^1_k((T^1_k)^*Q)\to T^1_kQ$ is locally given by
$T^1_k(\tau^*_Q)(q^i,p^A_i,(u_A)^i,(u_A)^B_i)=(q^i,(u_A)^i)$,
 we obtain that
\begin{equation}\label{sec}
(\K_A)^i=v_A^i\, .
\end{equation}
Then, writing in local coordinates  the expression (\ref{evo2})
$$
\ds\sum_{A=1}^k (\omega_0)_A(FL(v)) \left( \K_A(v),(FL)_*(v)
\left( \displaystyle\frac{\partial}{\partial q^i}\Big\vert_v
\right) \right)=dE_L \left( \displaystyle\frac{\partial}{\partial
q^i}\Big\vert_v \right)\, ,
$$
we obtain that
$$
\ds\sum_{A=1}^k \left( v^k_A \displaystyle\frac{\partial^2
L}{\partial q^i\partial v^k_A}(v) - (\K_A)^A_j(v) \right) =
\ds\sum_{A=1}^k v^k_A \displaystyle\frac{\partial^2 L}{\partial
q^i\partial v^k_A}(v) - \displaystyle\frac{\partial L}{\partial
q^i}(v)\, .
$$

Therefore
\begin{equation}\label{kl}
\ds\sum_{A=1}^k(\K_A)^A_j= (\K_1)^1_i + (\K_2)^2_i + \ldots +
(\K_k)^k_i=\displaystyle\frac{\partial L}{\partial q^i}\, ,
\end{equation}
which means that every field operator $\K$ is locally given by
$$
\K_A= v^i_A\left(\displaystyle\frac{\partial}{\partial q^i} \circ
FL \right) + (\K_A)^B_i
 \left(\displaystyle\frac{\partial}{\partial p^B_i} \circ
FL \right), \quad 1 \leq A \leq k.
$$
where the components $(\K_A)^B_i$ satisfy the identity (\ref{kl}).

Equations (\ref{sec}) and (\ref{kl}) lead us to define local
solutions in a neighborhood of each point of $T^1_kQ$ satisfying
conditions $1$, $2$ and $3$ in definition \ref{def61},
$$
\K_A= v^i_A\left(\displaystyle\frac{\partial}{\partial q^i} \circ
FL \right) +\displaystyle\frac{1}{k}
 \displaystyle\frac{\partial L}{\partial
q^i}\left(\displaystyle\frac{\partial}{\partial p^A_i} \circ FL
\right)\, , \quad 1 \leq A \leq k\, ,
$$
and, by using a partition of the unity, we obtain global
solutions.

\begin{definition}\label{22}
$ \psi: \rk \to T^1_kQ$ is an integral section of the field
operator $\K$ if
$$
T^1_k(FL) \circ \psi^{(1)} = \K\circ\psi \, .
$$
\end{definition}

 Definition \ref{22} means that, for every $t\in\rk$,
$$
\K_A(\psi(t))=(FL)_*(\psi(t))\left(\psi_*(t) \left(
\ds\frac{\partial}{\partial t^A}\Big\vert_{t}\right) \right) \, ,
\quad 1\leq A \leq k \, ,
$$
because
$$
(T^1_k(FL)\circ \psi^{(1)})(t)= T^1_k(FL)(j^1_0 \psi_t)=j^1_0( FL
\circ \psi_t) \, ,
$$
where $\psi_t(\bar{t})=\psi(t+\bar{t})$.
Thus, the following
diagram is commutative \vspace{2cm}
\begin{center}
\begin{picture}(200,80)(0,0)
\put(15,15){\makebox(0,0){$T^1_kQ$}} \put(125,15){\makebox(0,0){$
(T^1_k)^*Q$}}
 \put(60,20){$FL$}
\put(30,15){\vector(1,0){73}} \put(125,100){\vector(0,-1){75}}
\put(125,110){\makebox(0,0){$T^1_k((T^1_k)^*Q)$}}
\put(30,25){\vector(1,1){75}} \put(130,60){$\tau_{(T^1_k)^*Q}$}
\put(60 ,70 ){$\K$} \put(15,110) {\makebox(0,0){$T^1_k(T^1_kQ)$}}
\put(15,100){\vector(0,-1){75}} \put(40,110 ){\vector(1,0){52}}
\put(20,70){$\tau_{T^1_kQ}$}
 \put(50,120){$T^1_k(FL)$}
\put(-75,25){\vector(1,1){75}} \put(-60,70 ){$\psi^{(1)}$}
 \put(-80,15){\makebox(0,0){$\rk$}}
 \put(-35,20){\makebox(0,0){$\psi$}}
 \put(-70,15){\vector(1,0){70}}
\end{picture}
\end{center}

\subsection{Properties of the field operators related to the Lagrangian formalism}

In this section we study the properties of the field operator in
relation to the Lagrangian field equations. In particular, we
generalize the properties of the evolution operator in mechanics
given in equation (\ref{K2a}).

\begin{proposition}\label{p1}
Let $L:T^1_kQ \to \r$ be a Lagrangian.
 $\psi:\rk \to T^1_kQ$ is an integral section of $\K$
if, and only if, $\tau_Q \circ \psi: \rk \stackrel{\psi}{\to}
T^1_kQ \stackrel{\tau_Q}{\to} Q$ is a solution to the
Euler-Lagrange
  equations (\ref{lageq1}).
\end{proposition}

\proof
 If $\psi:\rk \to T^1_kQ$ \, is locally given by
$\psi(t)=(\psi^i(t),\psi^i_A(t))$, then from (\ref{locfl}) we
obtain that
\begin{equation}\label{kfi1}
\begin{array}{lcl}
(FL \circ \psi)_*(t)\left(\ds\frac{\partial}{\partial
t^A}\Big\vert_{t} \right) & = & \displaystyle\frac{\partial
\psi^j}{\partial t^A}(t) \, \,
\displaystyle\frac{\partial}{\partial q^j}\Big\vert_{FL(\psi(t))}
\\ \noalign{\medskip} & + &
 \left(\displaystyle\frac{\partial^2 L}{\partial q^i \partial
v^j_C}(\psi(t)) \, \displaystyle\frac{\partial \psi^i}{\partial
t^A}(t) + \displaystyle\frac{\partial\psi^i_B}{\partial t^A}(t) \,
\displaystyle\frac{\partial^2 L}{\partial v^i_B  \partial
v^j_C}(\psi(t)) \right)\, \displaystyle\frac{\partial}{\partial
p^C_j}\Big\vert_{FL(\psi(t))} \, .
\end{array}
\end{equation}

On the other hand
\begin{equation}\label{kfi2}
\K_A(\psi(t))= v^j_A(\psi(t))\,
\displaystyle\frac{\partial}{\partial q^j}\Big\vert_{FL(\psi(t))}
+ (\K_A)^C_j(\psi(t))\, \displaystyle\frac{\partial}{\partial
p^C_j}\Big\vert_{FL(\psi(t))} \, .
\end{equation}

So if $\psi$ is  a solution to $\K$, then from (\ref{kfi1}) and
(\ref{kfi2}) we obtain the equations
\begin{equation}\label{kfi3}
\displaystyle\frac{\partial \psi^j}{\partial t^A}(t)=
v^j_A(\psi(t))= \psi^j_A(t)\, ,
\end{equation}
and
\begin{equation}\label{kfi4}
\displaystyle\frac{\partial}{\partial t^A}\Big\vert_t
\left(\displaystyle\frac{\partial L}{\partial
v^j_C}(\psi(t))\right)= \displaystyle\frac{\partial^2 L}{\partial
q^i \partial v^j_C}(\psi(t)) \, \displaystyle\frac{\partial
\psi^i}{\partial t^A}(t) + \displaystyle\frac{\partial^2
\psi^i}{\partial t^A \partial t^B}(t) \,
\displaystyle\frac{\partial^2 L}{\partial v^i_B  \partial
v^j_C}(\psi(t)) = (\K_A)^C_j(\psi(t))\, ,
\end{equation}
 for every $A=1, \ldots , k$. Therefore, from (\ref{kl})
(\ref{kfi3}) and (\ref{kfi4}) we obtain
$$
\sum_{A=1}^k \displaystyle\frac{\partial}{\partial t^A}\Big\vert_t
\left(\displaystyle\frac{\partial L}{\partial
v^i_A}(\psi(t))\right)= \sum_{A=1}^k(\K_A)^A_j(\psi(t))
=\displaystyle\frac{\partial L}{\partial q^i}(\psi(t)) \, , \quad
\psi^i_A(t)= \displaystyle\frac{\partial \psi^i}{\partial
t^A}(t)\, ,
$$
that is  $(\tau_Q \circ \psi)(t)=(\psi^i(t))$ is a solution to the
Euler-Lagrange equations (\ref{lageq1}).

The proof of the converse follows the same pattern than in the
proof of the converse statement of proposition \ref{53}.
\begin{flushright}
 \blob \end{flushright}

\begin{theorem}\label{th61}
Let $L:T^1_k Q \to \r$ be a Lagrangian and let $\K$ be a $k$
vector field along the Legendre map $FL:T^1_k Q \to (T^1_k)^* Q$.
If ${\bf X}_L : T^1_kQ \to T^1_k(T^1_kQ)$ is a $k$-vector field on
$T^1_kQ$ and $j_S : S\hookrightarrow T^1_kQ$ is a submanifold of
$T^1_kQ$ such that
\begin{equation}
\label{K2} T^1_k(FL) \circ {\bf X}_L \feble{S} \K
\end{equation}
then $\K$ is a field operator associated with the Lagrangian $L$
if, and only if, ${\bf X}_L$ is a {\sc sopde} solution to the equation
(\ref{lageq0}).
$$
\begin{array}{ccc}
T^1_k(T^1_kQ) &
\begin{picture}(130,20)(0,0)
\put(53,8){\mbox{$T^1_k(FL)$}} \put(0,3){\vector(1,0){135}}
\end{picture}
& T^1_k((T^1_k)^*Q)
\\
\begin{picture}(15,35)(0,0)
\put(-13,15){\mbox{${\bf X}_L$}} \put(8,0){\vector(0,1){35}}
\end{picture}
 &
\begin{picture}(135,35)(0,0)
\put(45,19){\mbox{$\K$}} \put(0,0){\vector(4,1){135}}
\end{picture}
&
\begin{picture}(15,35)(0,0)
\put(3,15){\mbox{$\tau_{(T^1_k)^*Q}$}}
\put(0,35){\vector(0,-1){35}}
\end{picture}
\\
T^1_kQ &
\begin{picture}(135,10)(0,0)
\put(63,-7){\mbox{$FL$}} \put(0,7){\vector(1,0){135}}
\end{picture}
& (T^1_k)^*Q
\end{array}
$$
\label{equivlag}
\end{theorem}

\proof We must prove that both the second-order condition, and the
field equation condition hold for $\K$ if, and only if, they hold
for ${\bf X}_L$. In this proof all the equalities hold on $S$.

First, if $\K=(\K_1,\ldots,\K_k)$ and ${\bf
X}_L=((X_L)_1,\ldots,(X_L)_k)$, then equation (\ref{K2}) is
equivalent to
$$
T(FL)\circ (X_L)_A \feble{S} \K_A \, , \quad 1\leq A \leq k\, .
$$

On the other hand $(\omega_L)_A =(FL)^*(\omega_0)_A$ so one easily
proves that
$$
\imath_{(X_L)_A} \, (\omega_L)_A \feble{S} (FL)^*(\imath_{\K_A} \,
(\omega_0)_A \circ FL)\, ,$$ and for the field equation we obtain
$$
\sum_{A=1}^k[(FL)^*(\imath_{\K_A}(\omega_A\circ FL)]-dE_L
\feble{S} \sum_{A=1}^k[\imath_{X_A}(\omega_L)_A]-dE_L
$$
hence the field equation condition holds for $\K$ if, and only if,
the Lagrangian field equation holds for ${\bf X}_L$.

Furthermore, in relation to the second-order condition (see
Definition \ref{sode0}) we have that
$$
T^1_k(\tau_Q^*)\circ\K=Id_{T^1_kQ} \quad \Leftrightarrow\quad
T^1_k(\tau_Q^*)\circ T^1_k(FL)\circ{\bf X}_L=Id_{T^1_kQ} \quad
\Leftrightarrow\quad T^1_k(\tau_Q)\circ {\bf X}_L= Id_{T^1_kQ}
$$
because  $FL$ is a fiber preserving map,  that is $\tau_Q^*\circ
FL=\tau_Q$, and hence $T^1_k(\tau_Q^*)\circ
T^1_k(FL)=T^1_k(\tau_Q)$. Thus the last equality is equivalent to
(\ref{sodedef}),
%$T^1_k\tau\circ{\bf X}=Id_{T^1_kQ}$,
and so the second order conditions for $\K$ and ${\bf X}_L$ are
related. \begin{flushright}
 \blob \end{flushright}

Finally, as an immediate consequence of propositions \ref{53} and
\ref{p1}, and theorem \ref{th61}, we have:

\begin{corollary}\label{inteor}
Under the hypotheses of Theorem \ref{th61}, $\psi : \rk \to S
\subset T^1_kQ$ is an integral section of the field operator $\K$
if, and only if, it is an integral section of the {\sc sopde} ${\bf
X}_L$. (This means that $\K$ is integrable if, and only if, ${\bf
X}_L$ is integrable).

Moreover, every integral section $\psi : \rk \to S \subset T^1_kQ$
is an holonomic section.
\end{corollary}

\subsection{Properties of the field operators related to the Hamiltonian formalism}

Next we analyze the properties of the field operator in relation
to the Hamilton-de Donder-Weyl field equations, generalizing the
properties of the evolution operator in mechanics given in Eqs.
(\ref{K3}) and (\ref{K4}).

\begin{theorem}
Let $L$ be an almost-regular Lagrangian function, and $\K$ a field
operator associated with $L$. If there exist a $k$-vector field
${\bf X}_0\colon{\cal P}\to T^1_k{\cal P}$, and a submanifold
$\jmath_S\colon S\hookrightarrow T^1_kQ$, such that
\begin{equation}
T^1_k\jmath_0\circ{\bf X}_0\circ FL_0 \feble{S}\K  \quad ,
\label{elinduced2c} \end{equation} then ${\bf X}_0$ is a solution
to the equation (\ref{HEo}) on $P=FL_0(S)$.

 Conversely, if ${\bf X}_0$ is a $k$-vector field
solution to the equation (\ref{HEo}), then the above relation
defines a $k$-vector field $\K$ along $FL$, which satisfy
conditions 1 and 2 of Definition \ref{tildek}, on $S$, but not
condition 3 (second-order condition) necessarily.

If $L$ is a hyper-regular Lagrangian function, then the same
results hold (with $S=T^1_kQ$). But in addition, in the converse
statements the $k$-vector field $\K$ along $FL$ also satisfies the
second-order condition 3 of Definition \ref{def61}, and hence it
is a field operator for $L$. \label{equivham}
\end{theorem}
\proof Equation (\ref{elinduced2c}) means that
\begin{equation}\label{70}
(j_0)_*(FL_0(s))\left((X_0)_A(FL_0(s))\right)=\K_A(s)\, ,\quad
s\in S\, , \quad 1 \leq A \leq k \, .
\end{equation}

Then, since $j_0\circ FL_0= FL$  and
$(j_0)^*(\omega_0)_A=\omega^0_A$ we deduce,from  (\ref{70}) that
$$
(FL)^*(\imath_{\K_A}((\omega_0)_A\circ FL))\feble{S}
(FL_0)^*(\imath_{(X_0)_A}\omega_A^0)
$$
and since $(FL_0)^*H_0=E_L$ we obtain
$$
\sum_{A=1}^k(FL)^*(\imath_{\K_A}((\omega_0)_A\circ FL))-dE_L
\feble{S} (FL_0)^* \left(
\sum_{A=1}^k(\imath_{(X_0)_A}\omega_A^0)- dH_0 \right)
$$
where all the equalities hold on $S$. But, as $FL_0$ is a
submersion, we obtain that
$$
\sum_{A=1}^k(FL)^*(\imath_{\K_A}((\omega_0)_A\circ
FL))-dE_L\feble{S}0 \quad \Longleftrightarrow \quad
\sum_{A=1}^k(\imath_{(X_0)_A}\omega_A^0)-dH_0 \feble{P} 0
$$
hence the field equation condition holds for $\K$ on $S$ if, and
only if, the Hamiltonian field equation holds for $X_0$ on
$P=FL_0(S)$.

For hyper-regular systems, the proof of these properties is the
same, but taking into acount that now ${\cal P}=(T^1_k)^*Q$, and
$FL_0=FL$. In addition, the $k$-vector field ${\bf X}_0\equiv{\bf
X}$ is defined everywhere in $(T^1_k)^*Q$. Thus, the only addendum
is to prove that, if ${\bf X}$ is a solution to the equation
(\ref{HEo}), then its associated $k$-vector field along $FL$,
$\K$, satisfies the second-order condition.
 As ${\bf X}$ is a $k$-vector field in $(T^1_k)^*Q$,
by definition it is a section of $\tau_{(T^1_k)^*Q}$, thus
$\tau_{(T^1_k)^*Q}\circ{\bf X}=Id_{(T^1_k)^*Q}$. Then, taking into
account that $FL$ is a diffeomorphism, and that
(\ref{elinduced2c}) reduces to ${\bf X}\circ FL=\K$, we have
that
$$
T^1_k(\tau_Q^*)\circ\K=T^1_k(\tau_Q^*)\circ{\bf X}\circ FL=
FL^{-1}\circ\tau_{(T^1_k)^*Q}\circ{\bf X}\circ FL=Id_{T^1_kQ}
$$
which is the second-order condition for $\K$. \begin{flushright}
\blob \end{flushright}

Then assuming all these relations, we have:

\begin{theorem}
$\K$ is integrable if, and only if, ${\bf X}_0$ is integrable. In
particular:
\begin{enumerate}
\item Let $FL_S\colon S\to P$ be the restriction of $FL_0$ to $S$
(that is, $\jmath_P\circ FL_S=FL_0\circ\jmath_S$). If
$\ds\psi\colon \rk\stackrel{\psi_S}{\longrightarrow}S
\stackrel{\jmath_S}{\hookrightarrow}T^1_kQ$
 is an integral section of $\K$ on $S$, then
$\ds\psi_0\colon \rk\stackrel{\psi_P}{\longrightarrow}P
\stackrel{\jmath_P}{\hookrightarrow}{\cal P}$
 is an integral section of ${\bf X}_0$ on $P$,
where $\psi_P:=FL_S\circ\psi_S$. \item
 Conversely, if
$\ds\psi_0\colon \rk\stackrel{\psi_P}{\longrightarrow}P
\stackrel{\jmath_P}{\hookrightarrow}{\cal P}$ is an integral
section of ${\bf X_0}$ on $P$, then the section $\ds\psi\colon
\rk\stackrel{\psi_S}{\longrightarrow}S
\stackrel{\jmath_S}{\hookrightarrow}T^1_kQ$ is an integral section
of $\K$ on $S$, for every $\psi_S\colon \rk\to S\subseteq T^1_kQ$
such that $\psi_P=FL_S\circ\psi_S$.

The section $\psi_S$, and hence $\psi:=\jmath_S\circ\varphi_S$,
are holonomic if, and only if, $\K$ satisfies the second-order
condition (and hence it is a field operator).
\end{enumerate}
\label{inteorbis}
\end{theorem}
\proof If the system is almost-regular, consider the diagram
\begin{equation}
\begin{array}{ccccc}
&
\begin{picture}(80,25)(0,0)
\put(75,10){\mbox{$T^1_k(FL\circ\psi)$}}
\put(0,5){\vector(1,0){210}}
\end{picture}
& & &
\\
T^1_k \rk &
\begin{picture}(80,10)(0,0)
\put(3,7){\mbox{$T^1_k(FL_0\circ\psi)$}}
\put(0,3){\vector(1,0){80}}
\end{picture}
 & T^1_k{\cal P} &
\begin{picture}(80,10)(0,0)
\put(20,6){\mbox{$T^1_k\jmath_0$}} \put(0,3){\vector(1,0){80}}
\end{picture} &
T^1_k((T^1_k)^*Q)
\\
\begin{picture}(10,60)(0,0)
\put(8,25){\mbox{$\tau_{\rk}$}} \put(5,60){\vector(0,-1){60}}
\end{picture}
 & &
\begin{picture}(10,60)(0,0)
\put(-12,40){\mbox{${\bf X}_0$}} \put(5,0){\vector(0,1){55}}
\end{picture}
& &
\begin{picture}(10,60)(0,0)
\put(8,25){\mbox{$\tau_{(T^1_k)^*Q}$}}
\put(5,55){\vector(0,-1){55}}
\end{picture}
\\
 \rk &
\begin{picture}(80,10)(0,0)
\put(3,-4){\mbox{$\psi$}} \put(-10,6){\vector(1,0){37}}
\put(32,0){\mbox{$T^1_kQ$}} \put(70,-5){\mbox{$FL_0$}}
\put(65,6){\vector(1,0){27}} \put(150,40){\mbox{$\K$}}
\put(53,19){\vector(3,1){160}} \put(15,20){\mbox{$\psi_0$}}
\put(-10,13){\vector(1,0){100}} \put(-10,-9){\vector(2,-1){105}}
\end{picture}
 & {\cal P} &
\begin{picture}(80,10)(0,0)
\put(35,10){\mbox{$\jmath_0$}} \put(0,3){\vector(1,0){80}}
\end{picture}
 &
(T^1_k)^*Q
\\
&
\begin{picture}(80,22)(0,0)
\put(135,2){\mbox{$FL$}} \put(53,15){\vector(1,0){160}}
\end{picture}
 & & &
\\
  &
\begin{picture}(80,50)(0,0)
\put(-10,30){\mbox{$\psi_S$}} \put(-24,67){\vector(1,-1){50}}
\put(37,0){\mbox{$S$}} \put(27,33){\mbox{$\jmath_S$}}
\put(65,40){\mbox{$\psi_P$}} \put(40,15){\vector(0,1){58}}
\put(65,7){\mbox{$FL_S$}} \put(55,3){\vector(1,0){40}}
\end{picture}
 &
\begin{picture}(10,50)(0,0)
\put(3,0){\mbox{$P$}} \put(10,35){\mbox{$\jmath_P$}}
\put(6,15){\vector(0,1){45}}
\end{picture}
 & &
\end{array}
\label{diagfin}
\end{equation}
(where ${\bf X}_0$ denotes any extension of the $k$-vector field
solution on $P$ to ${\cal P}$).

\begin{enumerate}
\item   If $\psi$ is an integral section of $\K$ then
\begin{equation}\label{an0}
K_A(\psi(t))=(FL\circ\psi)_*(t) \left( \ds\frac{\partial}{\partial
t^A}\Big\vert_{t} \right) \, , \quad 1\leq A \leq k\, ,
\end{equation}
\noindent but $FL\circ \psi=j_0 \circ \psi_0$ because
$$
FL \circ \psi=FL\circ j_S\circ\psi_S=j_0 \circ j_P \circ FL_S
\circ\psi_S=j_0\circ j_P \circ \psi =j_0 \circ \psi_0 \, ,
$$
\noindent therefore (\ref{an0}) is equivalent to
\begin{equation}\label{an1}
\K_A (\psi(t))= (j_0)_*(\psi_0(t))\left((\psi_0)_*(t)
\left(\ds\frac{\partial}{\partial t^A}\Big\vert_{t}\right)
    \right) \, , \quad 1\leq A \leq k\, .
\end{equation}

 Furthermore, from (\ref{70}) and taking into account that
 $FL_0\circ \psi=\psi_0$, we have that
\begin{equation}\label{an2}
 \K_A (\psi(t))= (j_0)_*(FL_0(\psi(t)))(X_0)_A(FL_0(\psi(t)))\, =
 \,
 (j_0)_*(\psi_0(t))\left((X_0)_A(\psi_0(t))\right)  \, ,  \end{equation}
\noindent then, from (\ref{an1}) and (\ref{an2}), taking into
account that $\jmath_0$ is an imbedding, we deduce
$$
(\psi_0)_*(t)\left(\ds\frac{\partial}{\partial
t^A}\Big\vert_{t}\right)= (X_0)_A(\psi_0(t))\  \, , \quad 1\leq A
\leq k\, .
$$
Hence, $\psi_0$ is integral section of ${\bf X}_0$.

\item The converse is proved by reversing the above reasoning. In
addition, the sections $\psi_S$ and $\psi:=\jmath_S\circ\psi_S$
are holonomic if, and only if, they are integral sections of a
second-order $k$-vector field along the Legendre map.
\end{enumerate}
If the system is hyper-regular the proof is analogous, but taking
${\cal P}=(T^1_k)^*Q$ and $FL_0=FL$. \begin{flushright} \blob
\end{flushright}

It is important to point out that, if the integrability condition
holds only in a submanifold ${\cal I}\hookrightarrow S$, then
 Theorem \ref{inteorbis}
% and \ref{inteorbibis}
only holds on ${\cal I}$ and
$ FL ({\cal I})$ (which is assumed to be a submanifold of $P$).

Observe also that Theorem \ref{inteorbis}, together with Theorem
\ref{inteor}, establish the equivalence between the Lagrangian
and Hamiltonian formalisms.

\subsection*{Acknowledgments}

We acknowledge the financial support of {\sl Ministerio de Ciencia
y Tecnolog\'\i a}, BFM2002-03493 and the Re\-search Project
PGIDT01PXI20704PR of Xunta de Gali\-cia.
We thank Mr. Jeff Palmer for his assistance in preparing the English version
of the manuscript.


\begin{thebibliography}{99}

\bibitem{aw1} A. Awane: ``$k$-symplectic structures'',
{\sl J. Math. Phys.} {\bf 33} (1992) 4046-4052.

\bibitem{aw2} A. Awane: ``$G$-spaces $k$-symplectic homog\`enes'',
{\sl J. Geom. Phys.} {\bf 13} (1994) 139-157.

\bibitem{aw3} A. Awane, M. Goze: {\sl Pfaffian systems,
$k$-symplectic systems}. Kluwer Academic Publishers , Dordrecht
(2000).

\bibitem{BGPR-eblhf}
C. Batlle, J. Gomis, J.M. Pons, N. Rom\'an-Roy: ``Equivalence
between the Lagrangian and Hamiltonian formalism
  for constrained systems'',
{\sl J. Math. Phys.} {\bf 27} (1986) 2953-2962.

\bibitem{CL-92}
J.F. Cari\~nena, C. L\'opez: ``The time evolution operator for
higher-order singular Lagrangians'', {\sl J. Mod. Phys.} {\bf 7}
(1992) 2447-2468.

\bibitem{cgt}
R. S. Clark, D.S. Goel: ``On the geometry of an almost tangent
structure'', {\sl Tensor (N. S.)} {\bf 24} (1972) 243-252.

\bibitem{CMC}
J. Cort\'es, S. Mart\'\i nez, F. Cantrijn:
 ``Skinner-Rusk approach to time-dependent mechanics'',
{\sl Phys. Lett. A} {\bf 300} (2002) 250-258.

\bibitem{bar4}
A. Echeverr{\'\i}a-Enr{\'\i}quez, C. L\'{o}pez, J. Mar\'{i}n-Solano,
M.C. Mu\~{n}oz-Lecanda, N. Rom\'{a}n-Roy: ``Lagrangian-Hamiltonian unified
formalism for field theory''. {\sl J. Math. Phys.} {\bf 45}(1)
(2004) 360-380.

\bibitem{bar5}
A. Echeverr{\'\i}a-Enr{\'\i}quez,
 J. Mar\'{i}n-Solano, M.C. Mu\~{n}oz-Lecanda, N. Rom\'{a}n-Roy:
``On the construction of ${\cal K}$-operators in field theory as
sections along Legendre maps''. {\sl Acta Appl. Math.} {\bf 77}
(2003) 1-40.

\bibitem{e}
H.A. Eliopoulos: ``Structures presque tangents sur les vari\'{e}t\'{e}s
diff\'{e}rentiables'',
 {\sl C. R. Acad. Sci. Paris S\'er . I Math.} {\bf 255} (1962) 1563-1565.

\bibitem{FP-90}
C. Ferrario, A. Passerini: ``Symmetries and constants of motion
for constrained  Lagrangian systems: a presymplectic version of
the  Noether theorem'', {\sl J. Phys. A: Math. Gen.} {\bf 23}
(1990) 5061-5081.

\bibitem{GaP-00}
J.A. Garc\'\i a,  J.M. Pons: ``Rigid and gauge Noether symmetries
for constrained systems'', {\sl Int. J. Modern Phys. A} {\bf
15}(29) (2000) 4681-4721.

\bibitem{Sarda2}
G. Giachetta, L. Mangiarotti, G. Sardanashvily: {\sl New
Lagrangian and Hamiltonian Methods in Field Theory}, World
Scientific Pub. Co., Singapore (1997).

\bibitem{Sarda1}
G. Giachetta, L. Mangiarotti, G. Sardanashvily:
 ``Covariant Hamilton equations for field theory'', {\sl J. Phys. A}
  {\bf 32}(32)  (1999) 6629--6642.

\bibitem{GP-ggf}
X. Gr\`acia, J.M. Pons: ``A generalized geometric framework for
constrained systems'', {\sl Diff. Geom. Appl.} {\bf 2} (1992)
223-247.

\bibitem{GP-01}
X. Gr\`{a}cia, J.M. Pons: ``On an evolution operator connecting
Lagrangian and Hamiltonian formalisms''. {\sl Lett. Math. Phys.}
{\bf 17}(3)(1989) 175--180.

\bibitem{GP-92b}
X. Gr\`acia, J.M. Pons: ``A Hamiltonian approach to Lagrangian
Noether transformations'', {\sl J. Phys. A: Math. Gen.} {\bf 25}
(1992) 6357-6369.

\bibitem{GP-00}
X. Gr\`acia, J.M. Pons: ``Singular Lagrangians: some geometric
structures along the Legendre map'', {\sl J. Phys. A: Math. Gen.}
{\bf 34} (2001) 3047-3070.

\bibitem{grif1}
J. Grifone: ``Structure presque-tangente et connexions I'', {\sl
Ann. Inst. Fourier} {\bf 22} (1972) 287-334.

\bibitem{grif2}
J. Grifone: ``Structure presque-tangente et connexions II''. {\sl
Ann. Inst. Fourier} {\bf 22} (1972) 291-338.

\bibitem{gun}
C. G\"{u}nther: ``The polysymplectic Hamiltonian formalism in field
theory and calculus of variations I: The local case''. {\sl J.
Differential Geom.} {\bf 25} (1987) 23-53.

\bibitem{Ka-82}
K. Kamimura, ``Singular Lagrangians and constrained
Hamiltonian systems, generalized canonical formalism'', {\sl Nuovo
Cim. B} {\bf 69} (1982) 33-54.

\bibitem{Kana}
I. V. Kanatchikov:  ``Canonical structure of classical field
theory in the polymomentum phase space'', {\sl  Rep. Math. Phys.}
{\bf 41}(1) (1998) 49--90.

\bibitem{klein}
J. Klein: ``Espaces variationelles et m\'{e}canique'', {\sl  Ann.
Inst. Fourier} {\bf 12} (1962) 1-124.

\bibitem{LMM-2002}
M. de Le\'on, J.C. Marrero, D. Mart\'\i n de Diego: ``A new
geometrical setting for classical field theories'', {\sl Classical
and Quantum Integrability}. Banach Center Pub. {\bf 59}, Inst. of
Math., Polish Acad. Sci., Warsawa (2002) 189-209.

\bibitem{mt1}
M. de Le\'{o}n, I. M\'endez, M. Salgado: ``$p$-almost tangent
structures'', {\sl Rend.   Circ. Mat. Palermo} Serie II {\bf
XXXVII} (1988), 282-294.

\bibitem{mt2}
M. de Le\'{o}n, I. M\'{e}ndez, M. Salgado: ``Integrable $p$--almost
tangent structures and tangent bundles of
$p^1$-ve\-lo\-ci\-ties'', {\sl Acta Math. Hungar.} {\bf 58}(1-2)
(1991) 45-54.

\bibitem{mod1}
M. de Le\'{o}n; Eugenio Merino, Jos\'{e} A. Oubi\~{n}a, Paulo Rodrigues,
Modesto Salgado: ``Hamiltonian systems on $k$-cosymplectic
manifolds'', {\sl J. Math. Phys.} {\bf 39}(2) (1998) 876--893.

\bibitem{mod2}
M. de Le\'{o}n; Eugenio Merino,  Modesto Salgado: ``$k$-cosymplectic
manifolds and Lagrangian field theories'', {\sl J. Math. Phys.}
{\bf 42}(5) (2001) 2092--2104.

\bibitem{mor}
A. Morimoto: ``Liftings of some types of tensor fields and
connections to tangent $p^r$-velocities'',
 {\sl Nagoya Qath. J.} {\bf 40} (1970) 13-31.

\bibitem{fam}
F. Munteanu, A.M. Rey, M. Salgado: ``The G\"{u}nther's formalism in
classical field theory: momentum map and reduction'', {\sl J.
Math. Phys.} {bf 45}(5) (2004) 1730--1751.

\bibitem{McN} M. McLean; L. K. Norris: Covariant field theory
on frame bundles of fibered manifolds. J. Math. Phys. 41 (2000),
no. 10, 6808--6823.

\bibitem{No1} L.K. Norris:
{Generalized symplectic geometry on the frame bundle of a
manifold}, Lecture given at the {AMS Summer Research Institute on
Differential Geometry, 1990}, at U.C.L.A.

\bibitem{No2} L.K. Norris:
Generalized symplectic geometry on the frame bundle of a manifold,
{\sl Proc. Symp. Pure Math.}  {\bf 54}, Part 2 (Amer. Math. Soc.,
Providence RI, 1993), 435-465.

\bibitem{No3} L.K. Norris: Symplectic geometry on $T^*M$ derived from
$n$-symplectic geometry on $LM$. {\sl J.~Geom.\ Phys.} {\bf 13}
(1994), 51-78.

\bibitem{No4} L.K. Norris: Schouten-Nijenhuis Brackets.
{\sl J. Math.\ Phys.} {\bf  38} (1997), 2694-2709.

\bibitem{No5}  L. K. Norris: $n$-symplectic algebra of observables in
covariant Lagrangian field theory.  J. Math. Phys. 42  (2001), no.
10, 4827--4845.

\bibitem{PV-00}
F. Pugliese, A.M. Vinogradov: ``On the geometry of singular
Lagrangians'', {\sl J. Geom. Phys.} {\bf 35} (2000) 35-55,

\bibitem{PV-00b}
F. Pugliese, A.M. Vinogradov: ``Discontinuous trajectories of
Lagrangian systems
 with singular hypersurface'',
{\sl J. Math. Phys.} {\bf 42}(1) (2001) 309-329.

\bibitem{Sd-95}
G. Sardanashvily: {\sl Generalized Hamiltonian Formalism for Field
Theory. Constraint Systems}, World Scientific, Singapore (1995).

\bibitem{skinner2}
R. Skinner, R. Rusk: ``Generalized Hamiltonian dynamics. I.
Formulation on $T^*Q \oplus TQ$'', {\sl J. Math. Phys.} {\bf
24}(11) (1983) 2589--2594.

\bibitem{Tu-76}
W.M. Tulczyjew: ``Les sous-variet\'es lagrangiennes et la
dinamique hamiltonienne'', {\sl C.R. Acad. Sc. Paris} {\bf t 283A}
(1976) 15-18.


\end{thebibliography}
\end{document}